\documentclass[%
 reprint,
 aps,
 showpacs,
 amsmath,amssymb,
 dvipdfmx,
 longbibliograpy
]{revtex4-1}

\usepackage{graphicx}
\usepackage{dcolumn}
\usepackage{color}
\usepackage{bm}
\usepackage{here}
\usepackage{subfigure}
\usepackage{tabularx}
\newcolumntype{C}{>{\centering\arraybackslash}X}
\newcolumntype{L}{>{\raggedright\arraybackslash}X}
\newcolumntype{R}{>{\raggedleft\arraybackslash}X}
\usepackage{multirow}

\usepackage[
colorlinks=true, 
pdfstartview=FitV, 
linkcolor=blue, 
citecolor=magenta, 
urlcolor=blue, 
bookmarks=true,
bookmarksnumbered=true,
pdftitle={},
pdfauthor={}
]{hyperref}

\usepackage{amsthm}

\usepackage{comment}
\newcommand{\sectionprl}[1]{{\em #1}\/.---}

\begin{document}
\title{Emergence of surface long-range order under uniform shear flow}
\author{Hiroyoshi Nakano$^1$, Yuki Minami$^2$, Taiki Haga$^3$ and Shin-ichi Sasa$^4$}
\affiliation{$^1$ Department of Applied Physics and Physico-Informatics, Keio University, Kanagawa 223-8522, Japan}
\affiliation{$^2$ Department of Physics, Zhejiang University, Hangzhou 310027, China}
\affiliation{$^3$ Department of Physics and Electronics, Osaka Prefecture University, Osaka 599-8531, Japan}
\affiliation{$^4$ Department of Physics, Kyoto University, Kyoto 606-8502, Japan}

\date{\today}

\begin{abstract}
We study the two-dimensional surface long-range order in a non-equilibrium steady state under shear flow using the three-dimensional conserved $O(N)$ model. Whereas the correlation on the surface is enhanced by increasing interactions within the surface, the long-range order cannot be realized at equilibrium because of divergent thermal fluctuations associated with the low dimensionality of the surface. Here, the shear flow is applied parallel to the surface, on which the flow is set to zero. Despite the shear flow not affecting the order parameter on the surface directly, the fluctuations at the surface are strongly suppressed by the flow away from the surface, leading to the surface long-range order. We demonstrate these results through an exact analysis in the large-$N$ limit, where non-linear fluctuations are self-consistently treated.
\end{abstract}

\maketitle

\sectionprl{Introduction}
Because time-reversal symmetry of statistical properties is broken under non-equilibrium conditions, out-of-equilibrium systems exhibit anomalous fluctuations that have never been observed in equilibrium systems. The most familiar examples are conserved quantities exhibiting long-range correlations even far from critical points~\cite{kawasaki1973theory,ronis1982nonlinear,garrido1990long,dorfman1994generic,wada2003anomalous}. The behavior contrasts with equilibrium systems, for which the correlation length remains microscopic except near critical points. Another example of anomalous fluctuations is that external flow strongly suppresses long-wavelength fluctuations associated with equilibrium phase transitions, changing the universality class of critical phenomena~\cite{onuki1979nonequilibrium,katz1984nonequilibrium,cates1989role,bruinsma1992shear,bassler1995existence,haga2015nonequilibrium,nakano2021long,minami2021rainbow}. The suppression of fluctuations is also observed in the polar phase of active matter~\cite{vicsek1995novel,toner1995long,toner1998flocks,toner2012reanalysis,nishiguchi2017long,chate2020dry}. The purpose of this Letter is to propose a remarkable critical phenomenon arising from the suppression of long-wavelength fluctuations out of equilibrium.

\begin{figure}[b]
\begin{center}
\includegraphics[width=8.6cm]{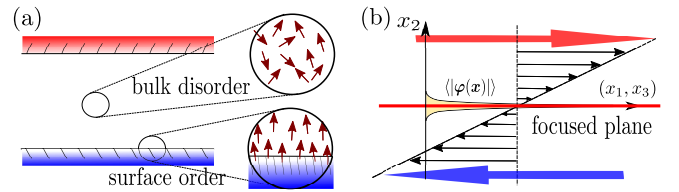}
\end{center}
\caption{Schematics of (a) surface long-range order, and (b) our setup. We study a specific plane at the center of an infinite system. The shear flow is applied parallel to the plane of focus. There is no net flow on the plane.}
\label{fig: schematic illustration of setup}
\end{figure}

The phenomenon under focus is the long-range order on surfaces~\cite{binder1983phase,diehl1986theory,dosch1992critical,diehl1997theory,pleimling2004critical}. We consider a system in which the order parameters move around between two parallel walls [see Fig.~\ref{fig: schematic illustration of setup} (a)]. Even though the bulk region far from the walls remains disordered, a localized long-range order may be realized near the walls, called surface long-range order. Numerous theoretical studies since the 1970s have been conducted on it in equilibrium systems such as ferromagnetic systems~\cite{mills1971surface,binder1972phase,binder1974surface,falicov1990surface}, fluids~\cite{nakanishi1982multicriticality,kellay1993prewetting,kozhevnikov1997prewetting,ancilotto1999prewetting}, nematic liquid crystals~\cite{bahr2009experimental}, block copolymer melts~\cite{fredrickson1987surface,stocker1996surface,mansky1997interfacial}, crystalline solids~\cite{lipowsky1983semi,frenken1985observation,dosch1991long}, and Bose--Einstein condensates~\cite{robinson1976bose,vandevenne2004equilibrium,nakano2019surface}. Because the presence or absence of surface long-range order can significantly change physical properties near the walls without changing the bulk properties, this phenomenon has recently attracted much attention in the fields of engineering and applied sciences~\cite{andrienko2003boundary,cottin2003low,seo2013situ,Lee2016}.

Surface long-range order is associated with critical phenomena in low-dimensional systems because the surface in the three-dimensional world is two-dimensional. A distinctive feature of low-dimensional systems is the absence of spontaneous breaking of continuous symmetries as described by the Hohenberg--Mermin--Wagner theorem~\cite{hohenberg1967existence,mermin1966absence,mermin1968crystalline}. It is then natural to ask whether the Hohenberg--Mermin--Wagner theorem can be applied to surface long-range order. This problem has been studied with various equilibrium models, such as the XY model~\cite{landau1989monte}, the Heisenberg model~\cite{wiener1973can,tsallis1988surface}, the $O(N)$ model~\cite{frohlich1986classical,deng2005surface,deng2006bulk}, and the spherical model~\cite{barber1974critical,singh1975critical}. To the best of our knowledge, there is no rigorous proof of the Hohenberg--Mermin--Wagner theorem for the surface long-range order. However, from the results of theoretical analyses and numerical simulations, the Hohenberg--Mermin--Wagner theorem is believed to hold for two-dimensional surfaces embedded in three-dimensional systems~\cite{diehl1997theory}.

In this Letter, we demonstrate through an exact analysis of the large-$N$-limit model that anomalous fluctuations in the non-equilibrium disordered bulk stabilize the two-dimensional surface long-range order prohibited in equilibrium. We study an infinite system with a specific two-dimensional plane where the interactions are enhanced [Fig.~\ref{fig: schematic illustration of setup} (b)]. The effect of the wall is modeled by changing the local interactions within the plane, and the long-range order localized on the plane is interpreted as the surface long-range order. We note that spatially localized order is one of the subjects studied in the context of surface critical phenomena~\cite{binder1983phase}. To drive the system into a non-equilibrium steady state, we apply uniform shear flow parallel to the plane. We consider the case in which the flow velocity on the plane is zero. Without knowing details about the surrounding bulk, one may consider this plane to be in equilibrium. We then ask whether spontaneous breaking of the continuous symmetry occurs on the plane.

\sectionprl{Model}
We consider an $N$-component real vector order parameter $\bm{\varphi}(\bm{x})=(\varphi^1(\bm{x}),\varphi^2(\bm{x}),\cdots,\varphi^N(\bm{x}))$ defined in a three-dimensional region $V \equiv [-L_1/2,L_1/2]\times[-L_2/2,L_2/2]\times [-L_3/2,L_3/2]$. The order parameter is convected by uniform shear flow with a velocity $\bm{v}(\bm{x}) = (\gamma x_2,0,0)$, conserving its total amount $\int_{V} d^3\bm{x} \varphi^a(\bm{x},t)$. We assume that the system is described by the time-dependent Ginzburg-Landau model~\cite{hohenberg1977theory}:
\begin{eqnarray}
\frac{\partial \varphi^a(\bm{x},t)}{\partial t} + \nabla \cdot \bm{j}^{a} = 0 ,
\label{eq:time dependent Ginzburg-Landau model} \\
\bm{j}^{a} = \varphi^a(\bm{x},t) \bm{v}(\bm{x}) - D_0 \nabla \biggl(\frac{\delta \Phi[\bm{\varphi}]}{\delta \varphi^a(\bm{x},t)} \biggr) + \bm{f}^a(\bm{x},t),
\label{eq:current definition}
\end{eqnarray}
where $\Phi[\bm{\varphi}]$ denotes the Landau free energy and $\bm{f}^a(\bm{x},t)$ Gaussian white noise satisfying $\big\langle \bm{f}^a(\bm{x},t) \big\rangle = \bm{0}$ and $\big\langle f_{\alpha}^{a}(\bm{x},t)f_{\beta}^{b}(\bm{x}',t') \big\rangle = 2D_0 T \delta_{\alpha \beta}\delta_{a b} \delta(\bm{x}-\bm{x}') \delta(t-t')$.
Here $T$ denotes the temperature of the thermal bath, and subscript $\alpha=1,2,3$ indices the spatial component. We also assume that $\int_{V} d^3\bm{x} \varphi^a(\bm{x},t) = 0$ for all $a$ throughout the dynamics.

The Landau free energy has two contributions, $\Phi[\bm{\varphi}] = \Phi_b[\bm{\varphi}] + \Phi_s[\bm{\varphi}]$. $\Phi_b[\bm{\varphi}]$ describes the standard $O(N)$ symmetric free energy and $\Phi_s[\bm{\varphi}]$ the local enhanced effects within the plane under focus. Each is given by~\cite{binder1983phase}
\begin{eqnarray}
& & \hspace{-1.5cm}\Phi_b[\bm{\varphi}] \nonumber \\
& & \hspace{-1.5cm} = \int d^3 \bm{x} \Big\{ \frac{1}{2}  \bm{\varphi}(\bm{x}) \cdot \Big(r_0 - \Delta \Big) \bm{\varphi}(\bm{x}) + \frac{g}{4N} (|\bm{\varphi}(\bm{x})|^2)^2 \Big\} ,
\label{eq:landau free energy in bulk}\\
\Phi_s[\bm{\varphi}] &=& \int d^{3} \bm{x}\frac{1}{2} \bm{\varphi}(\bm{x}) \cdot \Big(-c_0 \delta(x_2)\Big)\bm{\varphi}(\bm{x}).
\label{eq:landau free energy in surface}
\end{eqnarray}
Note that $r_0$ is chosen independently of the temperature $T$. In the Letter, we refer to the $x_2=0$ plane as the surface and the region sufficiently far from the $x_2=0$ plane as the bulk.

\begin{figure}[tb]
\begin{center}
\includegraphics[width=8.6cm]{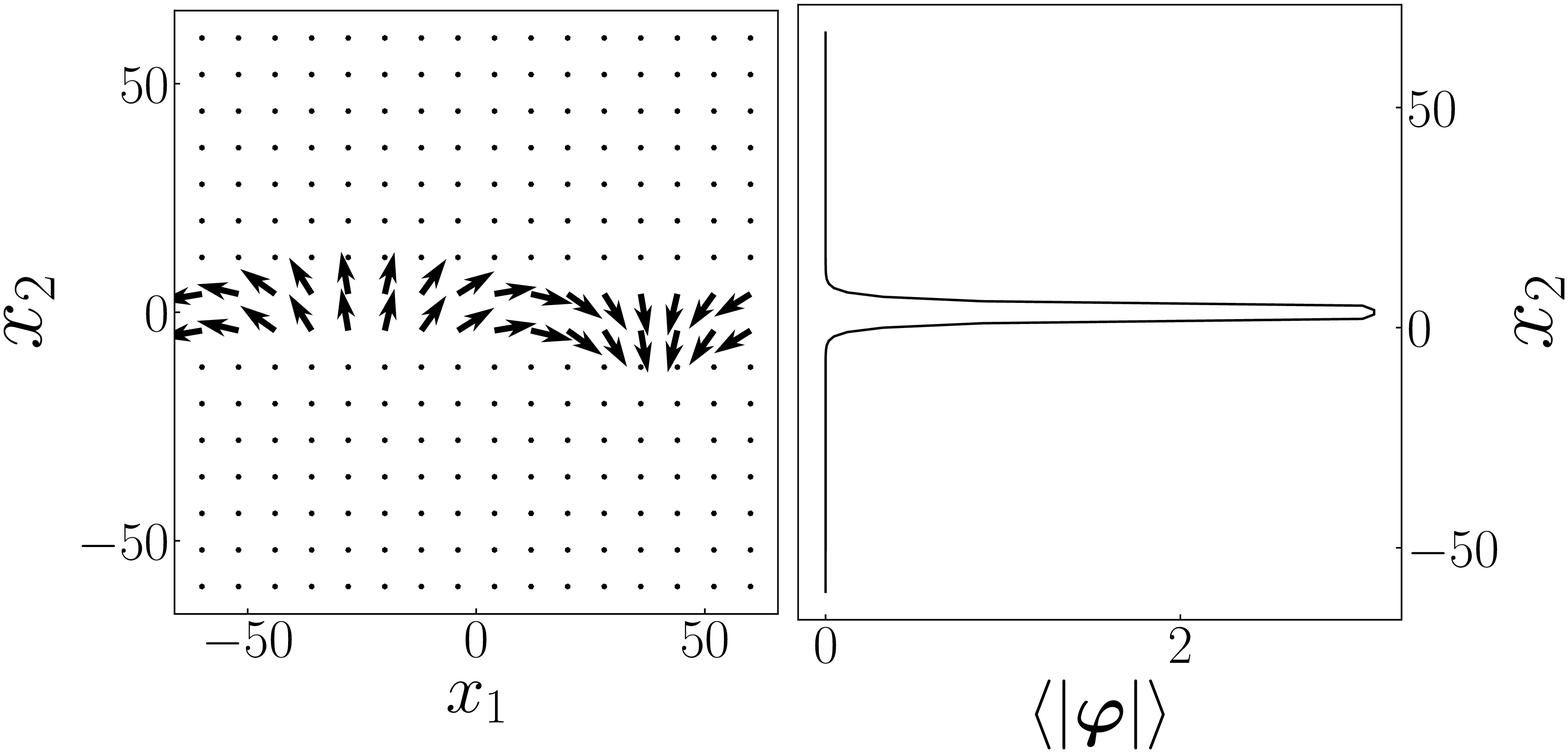}
\end{center}
\caption{Order parameter field at $T=0$ for $N=2$. Right: snapshot in the $x_2=0$ space. The order parameter is depicted as arrows. Left: amplitude of the order parameter.}
\label{fig: order parameter profile in ordered state}
\end{figure}

The point of interest is whether controlling $c_0$ stabilizes the two-dimensional long-range order at the surface while leaving the bulk disordered. We again stress that for any $\dot{\gamma}$ there is no flow at the surface, as explained in Introduction. To help in understanding the geometry of the surface long-range order, we evaluate the order parameter profile in the ordered state at $\dot{\gamma}=T=0$ (Fig.~\ref{fig: order parameter profile in ordered state}), which we obtained by numerically minimizing $\Phi[\bm{\varphi}]$ while taking into account the conservation law. The parameter settings are $N=2$, $r_0=g=1.0$, and $c_0=11.0$, and the system size is set at $L_1=L_2=L_3=128.0$. We use a mesh size of $1.0$ to discretize the space. The periodic behavior of $\bm{\varphi}$ near $x_2=0$ indicates the appearance of surface long-range order. Note that the conservation law prohibits the all-aligned state $\bm{\varphi}(\bm{x}_{\parallel},x_2=0)= (\varphi_0,0)$; here $\varphi_0$ is an appropriate constant~\cite{bassler1995existence,reichl2010phase}. Actually, the state shown in Fig.~\ref{fig: order parameter profile in ordered state} takes the form $\bm{\varphi}(\bm{x}_{\parallel},x_2=0)=\varphi_0\big(\cos(2\pi x_1/L_1),\sin(2\pi x_1/L_1)\big)$. The long-range order treated in this Letter is such a twisted state~%
\footnote{
See Supplemental Material for the phase diagram and the order parameter profile at $T=0$.
}. 

As evident in Fig.~\ref{fig: order parameter profile in ordered state}, the present model exhibits the localized ordered state at $T=0$. In the large-$N$ limit, the critical point $c_0^{\rm sc}(r_0;\dot{\gamma},T=0)$ is found to be independent of $\dot{\gamma}$ as
\begin{eqnarray}
\left. c_0^{\rm sc}(r_0;\dot{\gamma},T=0) = 1 \middle/ \int_{2\pi/L_2}^{2\pi/a^{\rm uv}_2} dk_2 \frac{1}{r_0+k_2^2} \right. ,
\label{eq:critical point at T=0}
\end{eqnarray}
where $a^{\rm uv}_{2}$ is the ultraviolet (UV) cutoff length of the $x_2$-axis. In the absence of shear flow, this ordered state is broken for any finite temperature $T>0$. However, by adding infinitesimal shear flow to that state, long-range order is restored. This is the main result, which we demonstrate below.

\sectionprl{Renormalization effect in the large-$N$ limit}
To perform an exact analysis of nonlinear effects, we consider the large-$N$ limit~\cite{amit1973field,amit1973self,corberi2003correlation}. In the following, $r_0$ is assumed to be sufficiently large so that the bulk remains disordered. When $c_0$ approaches the transition point $c_0^{\rm sc}(r_0;\dot{\gamma},T)$ from below, the disordered phase becomes unstable only near the surface. In other words, the correlation length at the surface diverges along the $x_1$- and $x_3$-axes but remains finite along the $x_2$-axis. We identify the transition point using this property without requiring the order parameter profile in the ordered state.

We define a new variable $\bm{\xi}$ to describe the local fluctuations of the order parameter,
\begin{eqnarray}
\bm{\xi}(\bm{x}) =  \bm{\varphi}(\bm{x}) - \big\langle \bm{\varphi}(\bm{x})\big\rangle.
\end{eqnarray}
When the whole area is disordered, the averaged order parameter profile is given by $\langle \bm{\varphi}(\bm{x})\rangle = \bm{0}$, and the time evolution of $\bm{\xi}$ is described by
\begin{eqnarray}
& & \hspace{-0.9cm} \Big[\frac{\partial }{\partial t} + \gamma x_2\frac{\partial}{\partial x_1} \Big]\xi^a(\bm{x},t) \nonumber \\
&=& D_0 \Delta \Big(-\Delta + r_0 - c_0\delta(x_2) +\frac{g}{N} |\bm{\xi}(\bm{x})|^2 \Big) \xi^a(\bm{x},t)  \nonumber \\
&-&  \nabla \cdot \bm{f}^a(\bm{x},t). 
\label{eq:equation of motion: xi}
\end{eqnarray}
The non-linear term of Eq.~(\ref{eq:equation of motion: xi}) is expanded in terms of $N$ as
\begin{eqnarray}
|\bm{\xi}(\bm{x})|^2\xi^a(\bm{x},t) = \big\langle |\bm{\xi}(\bm{x})|^2 \big\rangle \xi^a(\bm{x},t) + o(N),
\end{eqnarray}
where $o(N)$ represents the terms smaller than $N$. Equation~(\ref{eq:equation of motion: xi}) is then expressed in the large-$N$ limit in the linearized non-local form,
\begin{eqnarray}
& & \hspace{-1.2cm}\Big[\frac{\partial }{\partial t} + \gamma x_2\frac{\partial}{\partial x_1} \Big]\xi^a(\bm{x},t) \nonumber \\
&=& D_0 \Delta \big(-\Delta + r(r_0) - c(x_2;r_0,c_0) \big) \xi^a(\bm{x},t) \nonumber \\
&-&  \nabla \cdot \bm{f}^a(\bm{x},t),
\label{eq:equation of motion:large-N limit}
\end{eqnarray}
where $r(r_0)$ and $c(x_2;r_0,c_0)$ are, respectively, defined as
\begin{eqnarray}
r(r_0) &=& r_0 + \frac{g}{N} \lim_{x_2 \to \infty} \big\langle |\bm{\xi}(\bm{x})|^2 \big\rangle^{r,c()}_{l} ,
\label{eq:renormalization of r}\\[3pt]
c(x_2;r_0,c_0) &=& c_0 \delta(x_2)\nonumber \\
& & \hspace{-1.5cm} - \frac{g}{N}\Big(\big\langle |\bm{\xi}(\bm{x})|^2 \big\rangle^{r,c()}_{l} - \lim_{x_2 \to \infty} \big\langle |\bm{\xi}(\bm{x})|^2 \big\rangle^{r,c()}_{l} \Big). 
\label{eq:renormalization of c}
\end{eqnarray}
Here, to avoid confusion, we introduce the notation $\langle \cdot \rangle^{r,c()}_{l}$ as the average with respect to Eq.~(\ref{eq:equation of motion:large-N limit}), where $r$ and the functional form of $c$ are parameters contained in Eq.~(\ref{eq:equation of motion:large-N limit}). We note that $\lim_{x_2 \to \infty} \big\langle |\bm{\xi}(\bm{x})|^2 \big\rangle^{r,c()}_{l}$ is replaced with $\big\langle |\bm{\xi}(\bm{x})|^2 \big\rangle^{r,0}_{l}$ because the effect of the enhanced parameter $c_0$ decays exponentially from the surface. With this replacement, it becomes clear that the renormalized $r$ does not depend on $c_0$.

Because of the translational symmetry along the $x_1$- and $x_3$-axes in Eq.~(\ref{eq:equation of motion:large-N limit}), we rewrite $\big\langle |\bm{\xi}(\bm{x})|^2 \big\rangle^{r,c()}_{l}$ in terms of the Fourier transform $C_{l}^{r,c()}(\bm{k}_{\parallel},x_2,y_2;\dot{\gamma})$ defined by
\begin{eqnarray}
& & \hspace{-1.5cm} \frac{1}{N}\big\langle \bm{\xi}(\bm{x})\cdot \bm{\xi}(\bm{y})\big\rangle^{r,c()}_{l} \nonumber\\[3pt]
&=& \int \frac{d^2\bm{k}_{\parallel}}{(2\pi)^2} C_{l}^{r,c()}(\bm{k}_{\parallel},x_2,y_2;\dot{\gamma}) e^{i\bm{k}_{\parallel} \cdot (\bm{x}_{\parallel}-\bm{y}_{\parallel})}.
\label{eq:def of correlation function at surface}
\end{eqnarray}
Therefore, to solve Eq.~(\ref{eq:renormalization of c}) with respect to $c()$ in a self-consistent manner, we need to calculate $C_{l}^{r,c()}(\bm{k}_{\parallel},x_2,y_2;\dot{\gamma})$.
For this purpose, we assume that the functional form of $c$ is given by
\begin{eqnarray}
c(x_2)=c_0\delta(x_2) + c_1 e^{-x_2/\ell_1},
\label{eq:ansatz of c(x2)}
\end{eqnarray}
where $\ell_1$ is a constant of length dimension. Furthermore, we assume that $C^{r,c()}_{l}(\bm{k}_{\parallel},x_2,0;\dot{\gamma})$ also takes an exponential form
\begin{eqnarray}
C^{r,c()}_{l}(\bm{k}_{\parallel},x_2,0;\dot{\gamma}) = C^{r,c()}_{l}(\bm{k}_{\parallel},0,0;\dot{\gamma}) e^{-x_2/\ell_2[c()]},
\label{eq:ansatz of correlation length}
\end{eqnarray}
where $\ell_2[c()]$ depends on the functional form of $c$, but does not depend on $\bm{k}_{\parallel}$. These assumptions are reasonable because the correlation length along the $x_2$-axis remains finite even near the transition point.

Under these assumptions, we obtain~\footnote{%
See the Supplemental Material
}
\begin{eqnarray}
C_{l}^{r,c()}(\bm{k}_{\parallel},0,0,\dot{\gamma}) = \frac{TC_{l}^{r,0}(\bm{k}_{\parallel},0,0,\dot{\gamma})}{T - \bar{c} C_{l}^{r,0}(\bm{k}_{\parallel},0,0;\dot{\gamma})}
\label{eq:exact connection of two correlation functions: using ansatz}
\end{eqnarray}
with
\begin{eqnarray}
\bar{c} &=& c_0 + \ell_{\rm sum}c_1 ,
\label{eq:def of bar c}\\
\ell_{\rm sum} &=& \ell_1+\ell_2(0)+\ell_2[c()].
\end{eqnarray}
$C_{l}^{r,0}(\bm{k}_{\parallel},0,0;\dot{\gamma})$ is the correlation function for $c(x_2)=0$. Noting that Eq.~(\ref{eq:equation of motion:large-N limit}) with $c(x_2)=0$ describes the dynamics of $\bm{\xi}(\bm{x},t)$ at distances far from the $x_2=0$ plane, we interpret $C_{l}^{r,0}(\bm{k}_{\parallel},0,0;\dot{\gamma})$ as the correlation function in the disordered bulk. Accordingly, Eq.~(\ref{eq:exact connection of two correlation functions: using ansatz}) provides an expression for surface fluctuations in terms of the disordered bulk ones. This expression was already derived for $\dot{\gamma}=0$~\cite{bray1977critical}.

When $c(x_2)$ is set to $0$, Eq.~(\ref{eq:equation of motion:large-N limit}) recovers Galilean invariance~\cite{onuki1979nonequilibrium} and, as a result, $\big\langle \bm{\xi}(\bm{x})\cdot \bm{\xi}(\bm{y})\big\rangle^{r,0}_{l} = \big\langle \bm{\xi}(\bm{x}-\bm{y})\cdot \bm{\xi}(\bm{0})\big\rangle^{r,0}_{l}$ holds. We then introduce the Fourier transform $C^{r,0}_{l}(\bm{k};\dot{\gamma})$ defined by
\begin{eqnarray}
\frac{1}{N}\big\langle \bm{\xi}(\bm{x})\cdot \bm{\xi}(\bm{y})\big\rangle^{r,0}_{l} = \int \frac{d^3\bm{k}}{(2\pi)^3} C^{r,0}_{l}(\bm{k};\dot{\gamma}) e^{i\bm{k} \cdot (\bm{x}-\bm{y})}.
\end{eqnarray}
Noting the linearity of Eq.~(\ref{eq:equation of motion:large-N limit}), we derive the integral representation of $C^{r,0}_{l}(\bm{k};\dot{\gamma})$,~\cite{onuki1979nonequilibrium,nakano2021long}
\begin{eqnarray}
\hspace{-0.4cm} C^{r,0}_{l}(\bm{k};\dot{\gamma})= TD_0 \int_0^{\infty} ds e^{-\int_0^s d\lambda |\bm{\kappa}_{\lambda}|^2 ( |\bm{\kappa}_{\lambda}|^{2} + r)} |\bm{\kappa}_s|^{2}
\label{eq:formal expression of correlation function in the bulk}
\end{eqnarray}
with $\bm{\kappa}_\lambda = (k_1,k_2+\gamma \lambda k_1/2,k_3)$. Then, combining Eqs.~(\ref{eq:exact connection of two correlation functions: using ansatz}) and (\ref{eq:formal expression of correlation function in the bulk}), and numerically solving Eqs.~(\ref{eq:renormalization of r}) and (\ref{eq:renormalization of c}) with respect to $r$ and $c()$, we obtain $r$ and the functional form of $c$ as a function of $r_0$, $c_0$ and $T$.

\sectionprl{Transition point in the large-$N$ limit}
We calculate the transition point $c_0^{\rm sc}(r_0,\dot{\gamma},T)$ without solving Eqs.~(\ref{eq:renormalization of r}) and (\ref{eq:renormalization of c}) with respect to $r$ and $c()$. In terms of the correlation function on the surface, the transition point $c_0^{\rm sc}$ is identified as
\begin{eqnarray}
\lim_{\bm{k}_{\parallel} \to \bm{0}}C_{l}^{r,c()}(\bm{k}_{\parallel},0,0;\dot{\gamma}) \to \infty .
\label{eq:identified as the transition point}
\end{eqnarray}
Using Eq.~(\ref{eq:exact connection of two correlation functions: using ansatz}), this condition is immediately rewritten as $\bar{c} = c_{\rm max} \equiv T/C_{l}^{r,0}(\bm{k}_{\parallel}=\bm{0},0,0;\dot{\gamma})$. Then, the functional form of $c$ at the transition point $c_0=c_0^{\rm sc}$ is calculated from Eqs.~(\ref{eq:ansatz of c(x2)}) and (\ref{eq:def of bar c}), specifically
\begin{eqnarray}
c(x_2) = c^{\rm sc}_0 \delta(x_2) + \frac{c_{\rm max}-c^{\rm sc}_0}{\ell_{\rm sum}} e^{-x_2/\ell_1}.
\label{eq:c(x2) at critical point}
\end{eqnarray}
By applying Eq.~(\ref{eq:renormalization of c}) at the transition point, we obtain an expression for $c_0^{\rm sc}(r_0,\dot{\gamma},T)$, 
\begin{eqnarray}
c_0^{\rm sc}(r_0,\dot{\gamma},T) &=&  c_{\rm max} + g\ell_{\rm sum} \int_{2\pi/L}^{2\pi/a^{\rm uv}} \frac{d^{2}\bm{k}_{\parallel}}{(2\pi)^{2}} C^r_{\rm sc}(\bm{k}_{\parallel};\dot{\gamma}) \nonumber \\ 
&-& \ell_{\rm sum}(r-r_0),
\label{eq:expression of critical point}
\end{eqnarray}
where $a^{\rm uv}=a^{\rm uv}_1=a^{\rm uv}_3$ is the UV cutoff along the $x_1$- and $x_3$-axes, and $C^r_{\rm sc}(\bm{k}_{\parallel};\dot{\gamma})$ is the correlation function at the transition point given by
\begin{eqnarray}
\hspace{-0.8cm} C^r_{\rm sc}(\bm{k}_{\parallel};\dot{\gamma}) = \frac{C_{l}^{r,0}(\bm{k}_{\parallel},0,0,\dot{\gamma})}{1 - C_{l}^{r,0}(\bm{k}_{\parallel},0,0;\dot{\gamma})/C_{l}^{r,0}(\bm{k}_{\parallel}=\bm{0},0,0;\dot{\gamma})}.
\label{eq:correlation function of critical point}
\end{eqnarray}
Equation~(\ref{eq:critical point at T=0}) is derived immediately from Eq.~(\ref{eq:expression of critical point}) by considering the limit of $T\to +0$.

\sectionprl{Proof of existence of surface ordered state}
By noting that the surface long-range order occurs for $c_0>c_0^{\rm sc}(r_0,\dot{\gamma},T)$, we can judge its existence/non-existence from whether $c_0^{\rm sc}(r_0,\dot{\gamma},T)$ diverges to positive infinity. For example, for $T \to +0$, the surface ordered state exists as Eq.~(\ref{eq:critical point at T=0}) does not diverge. However, for $T>0$, $c_0^{\rm sc}(r_0,\dot{\gamma},T)$ may diverge due to the contribution from the infrared region in the wavenumber integral of Eq.~(\ref{eq:expression of critical point}). Therefore, we study an asymptotic expression of Eq.~(\ref{eq:correlation function of critical point}) for $|\bm{k}_{\parallel}| \ll r^{1/2}$.

\begin{figure}[t]
\begin{center}
\includegraphics[width=8.6cm]{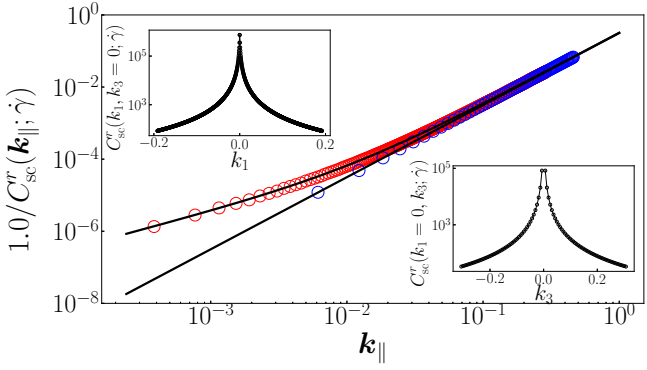}
\end{center}
\caption{Log-log plot of the inverse of $C^r_{\rm sc}(\bm{k}_{\parallel};\dot{\gamma})$. The red (blue) plot sets $k_3=0$ ($k_1=0$). The black line is Eq.~(\ref{eq:correlation function at surface: conserved}). The parameter settings are: $D_0=T=g=1$, $r=10$, $\dot{\gamma} = 0.3$, $L_1=16394$, and $L_2=L_3=1024$. The UV-cutoff length along the $x_2$-axis is chosen as $a^{\rm uv}_2= 0.01$. Inset: Linear-log plot of the same data.}
\label{fig: correlation function at surface: linearized model}
\end{figure}
The explicit expression for $C^r_{\rm sc}(\bm{k}_{\parallel};\dot{\gamma})$ is calculated by substituting Eq.~(\ref{eq:formal expression of correlation function in the bulk}) into Eq.~(\ref{eq:correlation function of critical point}). We present in Fig.~\ref{fig: correlation function at surface: linearized model} $C^r_{\rm sc}(\bm{k}_{\parallel};\dot{\gamma})$ obtained by numerical integration. We also derive an asymptotic expression of $C^r_{\rm sc}(\bm{k}_{\parallel};\dot{\gamma})$ for $|\bm{k}_{\parallel}| \ll r^{1/2}$ through rather complicated theoretical calculations~\footnote{%
See Supplemental Material for a detailed derivation
}; specifically, we have
\begin{eqnarray}
C^{r}_{\rm sc}(\bm{k}_{\parallel};\dot{\gamma}) \simeq \frac{T}{(2 \dot{\gamma}/\sqrt{3}D_0 r^{2})|k_1| + |\bm{k}_{\parallel}|^2/\sqrt{r}},
\label{eq:correlation function at surface: conserved}
\end{eqnarray}
where $a_2^{\rm uv} \ll r^{-1/2}$ is used. Equation (\ref{eq:correlation function at surface: conserved}) is plotted (Fig.~\ref{fig: correlation function at surface: linearized model}, black lines), and shows good agreement between the numerical and theoretical results.

We now substitute Eq.~(\ref{eq:correlation function at surface: conserved}) into Eq.~(\ref{eq:expression of critical point}) and study the convergence of the integral. For the equilibrium case, the wavenumber integration of Eq.~(\ref{eq:expression of critical point}) diverges logarithmically,
\begin{eqnarray}
\int_{2\pi/L}^{2\pi/a^{\rm uv}} \frac{d^{2}\bm{k}_{\parallel}}{(2\pi)^{2}} \frac{T}{|\bm{k}_{\parallel}|^2/\sqrt{r}} \sim \log L \to \infty
\label{eq:infrared divergence in equilibrium}
\end{eqnarray}
for $L =L_1=L_3 \to \infty$, which leads immediately to $c_0^{\rm sc}(r_0,\dot{\gamma}=0,T) = \infty$. Thus, we conclude that the two-dimensional surface long-range order does not appear in equilibrium.

In the presence of the shear flow, the situation is completely different. With the contribution of the $|k_1|$ term, Eq.~(\ref{eq:correlation function at surface: conserved}) is of order $L$ for $|\bm{k}|\sim2\pi/L$, which is considerably small compared with the terms of order $L^2$ in equilibrium. As a result, the logarithmic divergence of Eq.~(\ref{eq:infrared divergence in equilibrium}) is removed and the surface long-range order becomes stable, even in two dimensions. More specifically, when $\dot{\gamma}$ approaches $+0$, the infrared contribution of the wavenumber integration in Eq.~(\ref{eq:expression of critical point}) is cut off by $|\bm{k}|\simeq k_c \equiv 2\dot{\gamma}/(\sqrt{3}D_0 r^{3/2})$ instead of $2\pi/L$. $k_c$ is related to the length-scale governing the crossover behavior between $|k_x|^{-1}$ and $k_x^{-2}$ in $C^{r}_{\rm sc}(\bm{k}_{\parallel};\dot{\gamma})$. Then, the transition point works out to be
\begin{eqnarray}
& & c_0^{\rm sc}(r_0;\dot{\gamma},T)  \nonumber\\
&\simeq& 2\sqrt{r} + gl_{\rm sum} \int_{k_c}^{2\pi/a^{\rm uv}} \frac{d^{2}\bm{k}_{\parallel}}{(2\pi)^{2}} \frac{T}{|\bm{k}_{\parallel}|^2/\sqrt{r}} - \ell_{\rm sum} (r-r_0) \nonumber\\
&=& 2\sqrt{r} -\frac{g l_{\rm sum}T \sqrt{r}}{2\pi} \log \big(\frac{a^{\rm uv} \dot{\gamma}}{\sqrt{3}\pi D_0r^{3/2}}\big) - \ell_{\rm sum} (r-r_0). \nonumber \\[-8pt]
\label{eq:transition point: explicit form under shear}
\end{eqnarray}
We find that the transition point $c_0^{\rm sc}(r_0,\dot{\gamma},T)$ diverges as $-\log \dot{\gamma}$ for $\dot{\gamma} \to +0$. Conversely, for any nonzero $\dot{\gamma}$, the surface ordered state stabilizes for a sufficiently large $c_0$.

\begin{figure}[tb]
\begin{center}
\includegraphics[width=8.6cm]{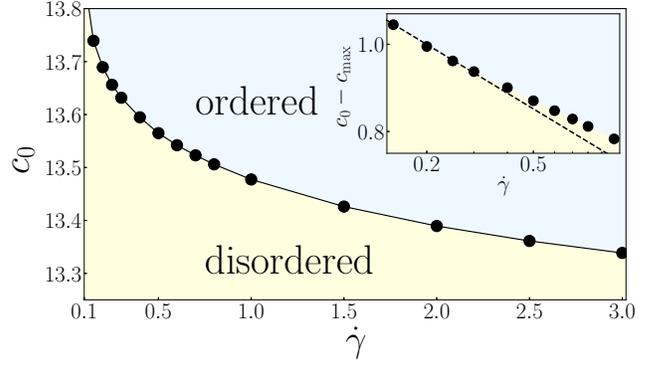}
\end{center}
\caption{Phase diagram in $(c_0,\dot{\gamma})$ space. Parameters: $D_0=T=g=l_{\rm sum}=1$, $r=10$, $L_1=16384$, and $L_2=L_3=1024$. The UV-cutoff lengths, $a^{\rm uv}_1$, $a^{\rm uv}_2$, and $a^{\rm uv}_3$ are set to $1.0$. Inset: the $\dot{\gamma}$-axis is logarithmically scaled. The black dashed line is a guideline given as $0.74-1.569 \log\dot{\gamma}$.}
\label{fig: phase diagram}
\end{figure}

We numerically calculate the transition point using Eqs.~(\ref{eq:formal expression of correlation function in the bulk}) and (\ref{eq:expression of critical point}). We neglect the renormalization effect of $r$ because it is quite small for sufficiently large $r_0$~%
\footnote{
see Supplemental Material for the renormalization of $r$.
}.
The results are plotted as a phase diagram in the $(c_0,\dot{\gamma})$-space (Fig.~\ref{fig: phase diagram}).
To extract the $\dot{\gamma}$-dependence of the transition point from the simulation data, we plot the line $0.74-1.569 \log\dot{\gamma}$ as a guide. The slope $-1.569$ is derived by considering the correction term for Eq.~(\ref{eq:correlation function at surface: conserved}) that comes from the condition $a_2^{\rm uv} \simeq r^{-1/2}$~%
\footnote{
See Supplemental Material for a detail argument of the correction term.
}. From this figure, we confirm that the transition point diverges as $\log\dot{\gamma}$ for $\dot{\gamma} \to +0$.

\sectionprl{Discussion}
We demonstrated that the uniform shear flow stabilizes the two-dimensional surface ordered state, which is prohibited in equilibrium. The interesting point is that the order parameter is not affected by the flow directly. Actually, the origin of this phenomenon is the anomalous fluctuations in the disordered non-equilibrium bulk. From Eq.~(\ref{eq:formal expression of correlation function in the bulk}), we calculate the asymptotic expression of the correlation function in this region as~\footnote{%
See Supplemental Materials for a detail derivation of the asymptotic expression.
}
\begin{eqnarray}
C^{r,0}_{l}(\bm{k};\dot{\gamma}) \simeq \frac{T}{r + |\bm{k}|^2 + A \{ (\dot{\gamma}/D_0 r)|k_1|\}^{2/3}},
\label{eq:correlation function in bulk: conserved}
\end{eqnarray}
where $A\simeq1.18$. We find that the fluctuations in the disordered bulk are suppressed as described by an unfamiliar term proportional to $|k_1|^{2/3}$. This term is present regardless of the state of the $x_2=0$ plane and leads to the $|k_1|$ mode in Eq.~(\ref{eq:correlation function at surface: conserved}). We note that by modifying the present model so as not to conserve the order parameter, the suppression of bulk fluctuations is removed, as a result, the localized two-dimensional order cannot be stabilized~%
\footnote{
See Supplemental Material for the dynamics where the order parameter is not conserved.
}.

The anomalous fluctuations in the disordered phase are one of the specific features of the non-equilibrium steady state. We expect the surface long-range order to be ubiquitous to these systems and may be accessed in experimental systems. One example is the isotropic-hexagonal order transition in a solution of diblock copolymers~\cite{cates1989role,koppi1992lamellae,koppi1993shear} or crystalline solids~\cite{elder2002modeling,elder2004modeling}, in which spatial translation symmetry is spontaneously broken while density is conserved. We also note that our model is related to the colloidal liquids made of nanoscale ferromagnetic particle, so-called ferrofluids~\cite{rosensweig2013ferrohydrodynamics,mrygold1995hydrodynamic,folk2000critical}. Ferrofluids exhibit a ferromagnetic phase transition with conserving the Heisenberg spin. Further exploration of the surface long-range order out of equilibrium is desirable.

\bigskip
\sectionprl{Acknowledgements}
We thank M. Hongo for stimulating conversations. H.N. was supported by KAKENHI Grant Number JP21J00034. Y.M. is supported by the Zhejiang Provincial Natural Science Foundation Key Project (Grant No. LZ19A050001) and NSF of China (Grants No. 11975199 and 11674283). T.H. was supported by KAKENHI Grant Number JP19J00525. S.S. was supported by KAKENHI Grant Numbers JP17H01148, JP19H05496, and JP19H05795.

%

\newpage
\setcounter{equation}{0}
\setcounter{figure}{0}
\setcounter{table}{0}
\setcounter{page}{1}
\renewcommand{\thepage}{S\arabic{page}}  
\renewcommand{\thesection}{S\arabic{section}}   
\renewcommand{\thetable}{S\arabic{table}}   
\renewcommand{\thefigure}{S\arabic{figure}}
\renewcommand{\theequation}{S\arabic{equation}}
\renewcommand{\bibnumfmt}[1]{[S#1]}
\renewcommand{\citenumfont}[1]{S#1}

\begin{widetext}
\begin{center}
{\large \bf Supplemental Material for  \protect \\ 
  ``Emergence of surface long-range order under uniform shear flow" }\\
\vspace*{0.3cm}
Hiroyoshi Nakano$^{1}$, Yuki Minami$^{2}$, Taiki Haga$^{3}$, and Shin-ichi Sasa$^{4}$
\\
\vspace*{0.1cm}
$^1${\small \it Department of Applied Physics and Physico-Informatics, Keio University, Kanagawa 223-8522, Japan} \\
$^2${\small \it Department of Physics, Zhejiang University, Hangzhou 310027, China}\\
$^3${\small \it Department of Physics and Electronics, Osaka Prefecture University, Osaka 599-8531, Japan}\\
$^4${\small \it Department of Physics, Kyoto University, Kyoto 606-8502, Japan}
\end{center}

\setcounter{equation}{0}
\renewcommand{\theequation}{S.\arabic{equation}}
\renewcommand{\thefigure}{S\arabic{figure}}
\renewcommand{\bibnumfmt}[1]{[S#1]}

\section{Phase diagram in equilibrium at zero temperature}
\label{sec:sup1}
We here give a detailed analysis regarding the properties of our model at $T=0$ in equilibrium. We treat both the cases of non-conserved and conserved order parameters. The contents of this section, especially for the case of non-conserved order parameter, were found in the literatures~\cite{sup:binder1983phase}. For simplicity, we restrict ourselves to the case $N=2$; the extension to $N>2$ can be straightforwardly performed.

\subsection{Bulk long-range order}
In this section, we study the phase diagram for regions sufficiently far from the $x_2=0$ plane. Because the effects of the enhanced interactions are neglected in this region, the phase diagram is calculated by minimizing the bulk free energy $\Phi_b[\bm{\varphi}]$. Then, we start with calculating local minima of $\Phi_b[\bm{\varphi}]$ by solving the following equation
\begin{eqnarray}
\frac{\delta \Phi_b[\bm{\varphi}]}{\delta \varphi^a(\bm{x},t)} = - \Delta \varphi^a(\bm{x}) + r_0 \varphi^a(\bm{x}) + \frac{g}{2} |\bm{\varphi}(\bm{x})|^2 \varphi^a(\bm{x}) = 0.
\label{eq:sup:the minimum of the total free energy: bulk}
\end{eqnarray}
We search the global-minimum state among the local-minimum solutions. It should be noted that the global-minimum state depends on the existence of the conservation law:
\begin{eqnarray}
\int_{V_3} d^3\bm{x} \bm{\varphi}(\bm{x}) = \bm{0}.
\label{eq:sup:conservation law}
\end{eqnarray}

First, we consider the non-conserved case. We choose a symmetry breaking solution by setting $\varphi^2(\bm{x})=0$. Then, Eq.~(\ref{eq:sup:the minimum of the total free energy: bulk}) becomes
\begin{eqnarray}
- \Delta \varphi^1(\bm{x}) + r_0 \varphi^1(\bm{x}) + \frac{g}{2} \big(\varphi^1(\bm{x})\big)^3 = 0.
\label{eq:sup:the minimum of the total free energy: bulk: mod 1}
\end{eqnarray}
The solution of Eq.~(\ref{eq:sup:the minimum of the total free energy: bulk: mod 1}) that satisfies the periodic boundary condition is given by
\begin{eqnarray}
\varphi^1(\bm{x}) = \left\{
\begin{array}{ll}
0 & {\rm for} \ r_0\geq 0, \\
\pm a_0 & {\rm for} \ r_0< 0, \\
\end{array}\right.
\label{eq:sup:solution of bulk equation: model A}
\end{eqnarray}
where we set
\begin{eqnarray}
a_0 = \sqrt{-\frac{2r_0}{g}} .
\end{eqnarray}
This result means that the bulk region exhibits the long-range order for $r_0<0$.  

Next, we consider the conserved case, where the solution Eq.~(\ref{eq:sup:solution of bulk equation: model A}) is not relevant because it does not satisfy the conservation law Eq.~(\ref{eq:sup:conservation law}). 
Instead of Eq.~(\ref{eq:sup:solution of bulk equation: model A}), we consider two types of solutions: domain-wall and twisted solutions. 
The domain-wall solution has a domain wall described by~\cite{sup:chaikin1995principles}
\begin{eqnarray}
\varphi^1(\bm{x}) = a_0 \tanh \Big(\frac{x_1}{\sqrt{2}\xi} \Big),
\end{eqnarray}
where we have assumed that the domain wall is located at $x_1=0$ and $\xi=r_0^{-1/2}$ is the correlation length. Then, the solution satisfying the conservation law Eq.~(\ref{eq:sup:conservation law}) is constructed by combining two domain walls, for example, as
\begin{eqnarray}
\varphi^1(\bm{x}) = a_0\tanh \Big(\frac{x_1+\frac{L_1}{4}}{\sqrt{2}\xi} \Big) - a_0\tanh \Big(\frac{x_1-\frac{L_1}{4}}{\sqrt{2}\xi} \Big) - a_0 + b(x_1),
\label{eq:sup:solution of bulk equation: model B}
\end{eqnarray}
where $b(x_1)$ is a small correction.
Next, in order to calculate the twisted solution, we return to Eq.~(\ref{eq:sup:the minimum of the total free energy: bulk}). We here assume that $|\bm{\varphi}(\bm{x})|^2$ is a constant $I_b$ independent of $\bm{x}$:
\begin{eqnarray}
|\bm{\varphi}(\bm{x})|^2 = I_b,
\label{eq:sup:assumption for bulk solution O(2) to O(1)}
\end{eqnarray}
and also assume that $\bm{\varphi}(\bm{x})$ depends only on the $x_1$-coordinate. Then, Eq.~(\ref{eq:sup:the minimum of the total free energy: bulk}) is simplified as
\begin{eqnarray}
 - \frac{\partial^2}{\partial x^2_1}\varphi^a(\bm{x}) + \big(r_0+ \frac{g}{2} I_b\big) \varphi^a(\bm{x}) = 0,
\end{eqnarray}
and the general solution is immediately obtained as
\begin{eqnarray}
\varphi^a(\bm{x}) &=& A_1^a e^{\sqrt{r_0+gI_b/2} x_1} + A_2^a e^{-\sqrt{r_0+gI_b/2} x_1}.
\end{eqnarray}
By imposing the periodic boundary condition, $I_b$ is calculated as
\begin{eqnarray}
I_b = \left\{
\begin{array}{ll}
0 & {\rm for} \ r_0 \geq 0, \\
-\frac{2}{g} \big(r_0 + \frac{(2n\pi)^2}{L_1^2} \big) & {\rm for} \ r_0 < 0,
\end{array}\right.
\end{eqnarray}
where $n=1,2,\cdots$. The constants $(A_1^a,A_2^a)$ are determined by the condition Eq.~(\ref{eq:sup:assumption for bulk solution O(2) to O(1)}). The final expression of twisted solutions is given by
\begin{eqnarray}
\left(\begin{array}{ll}
\varphi^1(\bm{x}), & \varphi^2(\bm{x})
\end{array}\right)
= \left(
\begin{array}{ll}
a_n\cos\big(\frac{2n\pi}{L_1} x_1\big), & a_n\sin\big(\frac{2n\pi}{L_1} x_1\big)
\end{array}\right)
\label{eq:sup:solution of bulk equation: O(0)}
\end{eqnarray}
with
\begin{eqnarray}
a_n = \sqrt{-\frac{2r_0}{g}-\frac{2(2n\pi)^2}{gL_1^2}}
\end{eqnarray}
for $r_0<0$ and $\varphi^1(\bm{x})=\varphi^2(\bm{x})=0$ for $r_0\geq0$.
We here notice that the total free energy $\Phi_b[\bm{\varphi}]$ increases with increasing $n$. Therefore, we have the twisted solution with $n=1$ as a candidate of minimizing the total free energy $\Phi_b[\bm{\varphi}]$.

From the above calculation, we find that the bulk long-range order occurs for $r_0<0$ regardless of the presence or absence of conservation law. For the non-conserved case, it is trivial that Eq.~(\ref{eq:sup:solution of bulk equation: model A}) gives the global minimum of $\Phi_b[\bm{\varphi}]$. In contrast, for the conserved case, we need to compare $\Phi_b[\varphi]$ for the two solutions, the domain-wall and twisted solutions. For the domain-wall solution, the domain wall gives the extra free energy that is proportional to its area $O(L_2 L_3)$. In contrast, for the twisted solution with $n=1$, the extra free energy comes from the spatial variation of the order parameter, which is estimated as $O(L_2 L_3/L_1)$. Therefore, when the system size is sufficiently large, the twisted solution Eq.~(\ref{eq:sup:solution of bulk equation: O(0)}) with $n=1$ is realized in the ordered phase.

\subsection{Surface long-range order}
We study the surface long-range order for $c_0>0$. For this purpose, $r_0$ is chosen to be positive ($r_0>0$) so that the regions sufficiently far from the $x_2=0$ plane remain disordered. We also assume that the ultraviolet cutoff in the $x_2$-axis, $a_2^{\rm uv}$, is sufficiently smaller than $r_0^{-1/2}$. We show that the surface long-range order occurs in equilibrium at $T=0$. Here, the surface long-range order is identified by two properties; (i) the long-range order on the $x_2=0$ plane and (ii) the exponential decay into the bulk.

The state that minimizes the total free energy $\Phi_b[\bm{\varphi}]+\Phi_s[\bm{\varphi}]$ is given by~\cite{sup:lubensky1975critical}
\begin{eqnarray}
- \Delta \varphi^a(\bm{x}) + r_0 \varphi^a(\bm{x}) + \frac{g}{2} |\bm{\varphi}(\bm{x})|^2 \varphi^a(\bm{x}) = 0
\label{eq:sup:zero temperature theory:bulk}
\end{eqnarray}
with the following boundary conditions on the $x_2=0$ plane
\begin{eqnarray}
\frac{\partial\bm{\varphi}(+0)}{\partial x_2} = - \frac{c_0}{2}\bm{\varphi}(+0),
\label{eq:sup:zero temperature theory:boundary 1}\\[3pt]
\frac{\partial \bm{\varphi}(-0)}{\partial x_2}  = \frac{c_0}{2} \bm{\varphi}(-0).
\label{eq:sup:zero temperature theory:boundary 2}
\end{eqnarray}
Here, we note that the additional free energy $\Phi_s[\bm{\varphi}]$ yields the boundary conditions at $x_2=0$. Because we are interested in the regime where the bulk is still disordered, we impose two additional boundary conditions:
\begin{eqnarray}
 \bm{\varphi}(\pm \infty) = \bm{0}, 
\label{eq:sup:zero temperature theory:boundary 3} \\[3pt]
\frac{\partial \bm{\varphi}(\pm \infty) }{\partial x_2}= \bm{0},
\label{eq:sup:zero temperature theory:boundary 4}
\end{eqnarray}
where we have taken the large system-size limit $L_2\to \infty$. These boundary conditions mean that the bulk remains disordered. We then assume that $\big|\bm{\varphi}(\bm{x})\big|^2$ depends only on the $x_2$-coordinate:
\begin{eqnarray}
|\bm{\varphi}(\bm{x})|^2 = I(x_2).
\label{eq:sup:assumption of zero temperature theory: mod}
\end{eqnarray}
This assumption corresponds to Eq.~(\ref{eq:sup:assumption for bulk solution O(2) to O(1)}) in the bulk long-range order.

By multiplying Eq.~(\ref{eq:sup:zero temperature theory:bulk}) by $\partial \varphi^a /\partial x_2$, we obtain
\begin{eqnarray}
\hspace{-1cm}-\frac{1}{2}\frac{\partial^2}{\partial x_1^2}\Big(\frac{\partial}{\partial x_2}(\varphi^a)^2\Big) - \frac{1}{2}\frac{\partial}{\partial x_2}\Big(\frac{\partial \varphi^a}{\partial x_2}\Big)^2 + \frac{1}{2}r_0 \frac{\partial}{\partial x_2}(\varphi^a)^2 +  \frac{g}{2} I(x_2) \frac{\partial}{\partial x_2}(\varphi^a)^2 = 0.
\end{eqnarray}
By taking the sum with respect to $a$ and integrating from $x_2=+0$ to $x_2=\infty$, $I(x_2)$ at the $x_2=0$ plane is calculated as
\begin{eqnarray}
\frac{1}{2}\biggl(\frac{c_0^2}{4}-r_0\biggr) I(+0) - \frac{g}{4} I(+0)^2  = 0.
\label{eq:sup:zero temperature theory:full mod}
\end{eqnarray}
From this equation, we obtain
\begin{eqnarray}
I(+0) = \left\{
\begin{array}{ll}
0 & {\rm for} \ c_0\leq2\sqrt{r_0}, \\
\frac{2}{g}\biggl(\frac{c_0^2}{4}-r_0\biggr) & \mbox{for} \ c_0>2\sqrt{r_0}. \\
\end{array}\right.
\label{surface order1}
\end{eqnarray}
Accordingly, the phase transition on the $x_2=0$ plane is observed at $c_0=2\sqrt{r_0}$. We note that this result coincides with that in the large-$N$ limit, Eq.~(5) if the ultraviolet cutoff $a_2^{\rm uv}$ is taken to be infinitesimal~\cite{sup:lubensky1975critical}.

In order to show that the order is localized near the $x_2=0$ plane, we return to Eq.~(\ref{eq:sup:zero temperature theory:bulk}). Because the boundary condition Eq.~(\ref{eq:sup:zero temperature theory:boundary 3}) implies that the order parameter is so small sufficiently far from the $x_2=0$ plane, we can neglect the non-linear term from Eq.~(\ref{eq:sup:zero temperature theory:bulk}) there. Then, Eq.~(\ref{eq:sup:zero temperature theory:bulk}) is rewritten as
\begin{eqnarray}
- \frac{\partial^2}{\partial x_2^2} \varphi^a(\bm{x}) + r_0 \varphi^a(\bm{x}) = 0,
\end{eqnarray}
and we immediately find that $\varphi^a(\bm{x})$ decays to $0$ with the correlation length $\xi=r_0^{-1/2}$. By combining this fact with Eq.~(\ref{surface order1}), we conclude that the surface long-range order is realized at $T=0$.

We then consider the order parameter profile in the $x_2=0$ plane. It is understood from the similar argument as the bulk long-range order. For the non-conserved case, because there is no constraint for the order parameter, the order parameter points to the same direction in the true equilibrium state, specifically, described by
\begin{eqnarray}
\left(\begin{array}{ll}
\varphi^1(x_1,x_2=0,x_3), & \varphi^2(x_1,x_2=0,x_3)
\end{array}\right)
= \left(
\begin{array}{ll}
\sqrt{\frac{2}{g}\big(\frac{c_0^2}{4}-r_0\big)},  & 0
\end{array}\right),
\end{eqnarray}
where the direction of order is assumed to be parallel to the $\varphi^1$-direction. 
In contrast, for the conserved case, we need to compare the domain-wall and twisted solutions.
For the domain-wall solution, the direction of order is fixed (e.g. $\varphi^1$) and the magnitude of the order parameter is given by $\sqrt{I(+0)}$.
Therefore, in order to satisfy the conservation law, there must be order-parameter flips somewhere in the $x_2=0$ plane (c.f. Eq.~(\ref{eq:sup:solution of bulk equation: model B})), which yield the extra energy proportional to $O(L_3)$. 
For the twisted solution, the order-parameter profile in the $x_2=0$ plane is given by
\begin{eqnarray}
\left(\begin{array}{ll}
\varphi^1(x_1,x_2=0,x_3), & \varphi^2(x_1,x_2=0,x_3)
\end{array}\right)
= \left(
\begin{array}{ll}
\sqrt{\frac{2}{g}\big(\frac{c_0^2}{4}-r_0\big)}\cos\big(\frac{2n\pi}{L_1} x_1\big), & \sqrt{\frac{2}{g}\big(\frac{c_0^2}{4}-r_0\big)}\sin\big(\frac{2n\pi}{L_1} x_1\big)
\end{array}\right).
\label{eq:sup:solution of surface equation: O(0)}
\end{eqnarray}
Then, we find that the spatial variation of the order parameter yields the extra energy proportional to $O(L_3/L_1)$. By comparing these two solutions, we conclude that the global minimum is given by Eq.~(\ref{eq:sup:solution of surface equation: O(0)}). This result corresponds to Fig.~2 in the main text.

In summary, we present the phase diagram in Fig.~\ref{fig:Phase diagram at zero temperature}, which is often found in the literatures~\cite{sup:binder1983phase}.
\begin{figure}[h]
\begin{center}
\includegraphics[width=8.6cm]{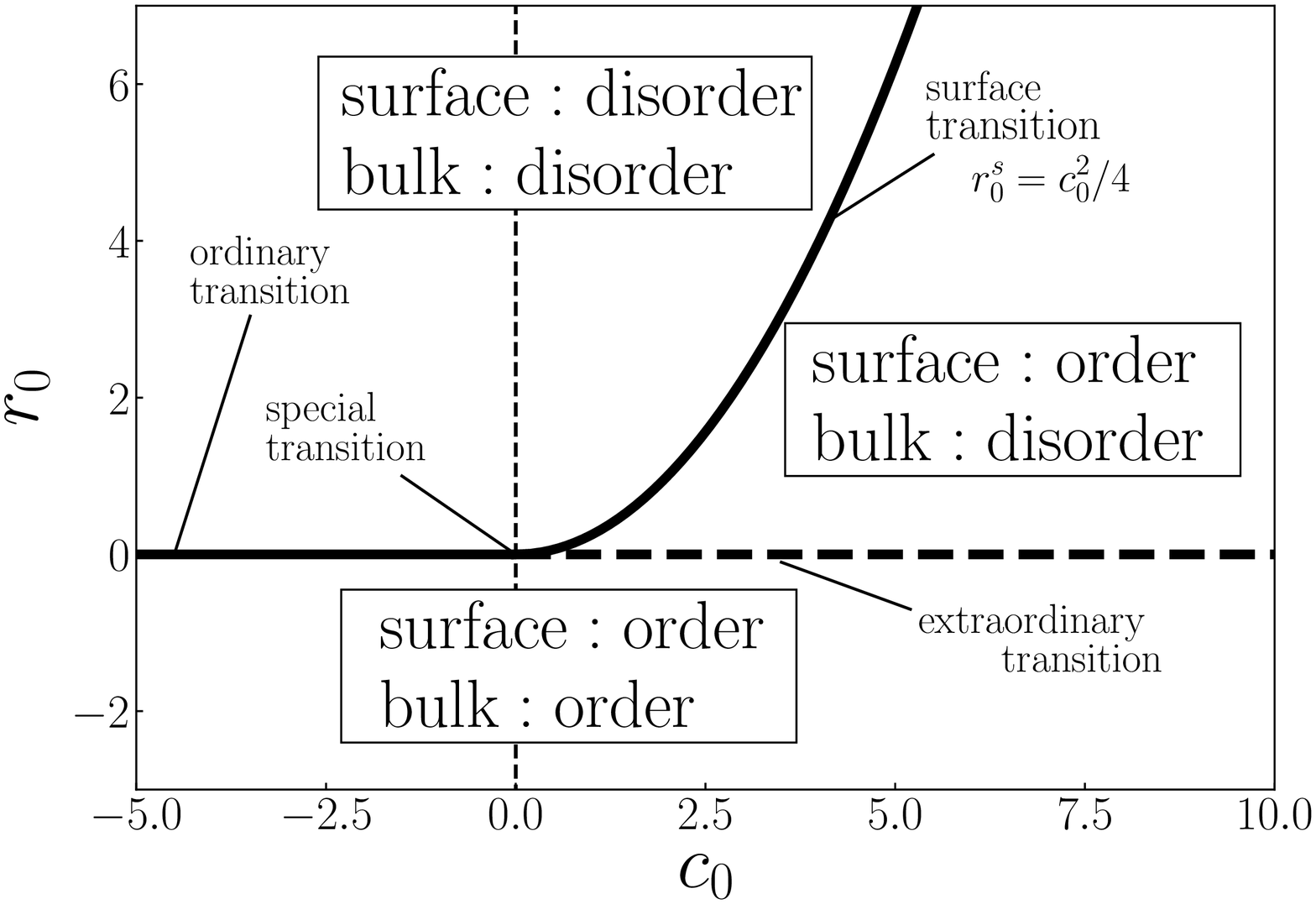}
\end{center}
\vspace{-0.5cm}
\caption{Phase diagram of our model at $T=0$ in equilibrium.}
\label{fig:Phase diagram at zero temperature}
\end{figure}

\section{Detailed analysis of linear fluctuations}
\label{sec:sup2}
In the main text, we used the results obtained by analyzing the linearized equation:
\begin{eqnarray}
\Big(\frac{\partial }{\partial t} + \dot{\gamma} x_2\frac{\partial}{\partial x_1} \Big)\xi^a(\bm{x},t) = D_0 \Delta \big(-\Delta + r - c(x_2)  \big) \xi^a(\bm{x},t)  - \nabla \cdot \bm{f}^a(\bm{x},t). 
\label{eq:sup:linearized equation of motion: appendix}
\end{eqnarray}
We here give their derivations.

The basic quantity of interest is the equal-time correlation function in the steady state, which is defined by
\begin{eqnarray}
C^{r,c()}_{l}(\bm{x},\bm{y};\dot{\gamma}) = \lim_{t \to \infty} C^{r,c()}_{l}(\bm{x},\bm{y},t;\dot{\gamma})
\end{eqnarray}
with
\begin{eqnarray}
C^{r,c()}_{l}(\bm{x},\bm{y},t;\dot{\gamma}) = \frac{1}{N}\big\langle\bm{\xi}(\bm{x},t)\cdot \bm{\xi}(\bm{y},t) \big\rangle^{r,c()}_{l} .
\end{eqnarray}
When the focus is restricted to the steady state with $c(x_2)=0$, the correlation function satisfies the translational invariance
\begin{eqnarray}
C^{r,0}_{l}(\bm{x},\bm{y};\dot{\gamma}) = C^{r,0}_{l}(\bm{x}+\bm{a},\bm{y}+\bm{a};\dot{\gamma}),
\label{eq:sup:translational invariance for correlation function: c=0}
\end{eqnarray}
where $\bm{a}$ is any vector.
This result is derived directly from Galilean invariance of Eq.~(\ref{eq:sup:linearized equation of motion: appendix}) with $c(x_2)=0$. The detailed discussion was given in Ref.~\cite{sup:onuki1979nonequilibrium}. Based on this property, we introduce the Fourier transform $C^{r,0}_{l}(\bm{k};\dot{\gamma})$ with respect to all the directions as
\begin{eqnarray}
C^{r,0}_{l}(\bm{x},\bm{y};\dot{\gamma}) =\int \frac{d^3\bm{k}}{(2\pi)^3}C^{r,0}_{l}(\bm{k};\dot{\gamma})e^{i\bm{k}\cdot (\bm{x}-\bm{y})}.
\end{eqnarray}

While Eq.~(\ref{eq:sup:translational invariance for correlation function: c=0}) does not hold for $c(x_2)\neq 0$, there remains the translational symmetry along the $x_1$- and $x_3$-directions. Then, we introduce the Fourier transform with respect to the remaining directions:
\begin{eqnarray}
C^{r,c()}_{l}(\bm{x},\bm{y};\dot{\gamma}) =\int \frac{d^2\bm{k}_{\parallel}}{(2\pi)^2}C^{r,c()}_{l}(\bm{k}_{\parallel},x_2,y_2;\dot{\gamma})e^{i\bm{k}_{\parallel}\cdot (\bm{x}_{\parallel}-\bm{y}_{\parallel})},
\label{eq:sup:Fourier transform w.r.t. 13direction}
\end{eqnarray}
where $\bm{x}_{\parallel}=(x_1,x_3)$ and $\bm{k}_{\parallel}=(k_1,k_3)$.

\subsection{Derivation of Eq.~(15)}
The equation for $C^{r,c()}_{l}(\bm{x},\bm{y};\dot{\gamma})$ is derived from Eq.~(\ref{eq:sup:linearized equation of motion: appendix}) as
\begin{eqnarray}
\Big\{\dot{\gamma} x_2\frac{\partial}{\partial x_1} + \dot{\gamma} y_2\frac{\partial}{\partial y_1} + D_0 \Delta_x\Big(\Delta_x - r + c(x_2) \Big) + D_0 \Delta_y\Big(\Delta_y - r + c(y_2)\Big) \Big\} C^{r,c()}_{l}(\bm{x},\bm{y};\dot{\gamma}) = -D_0 T \big(\Delta_x + \Delta_y\big)\delta(\bm{x}-\bm{y}), \nonumber \\
\label{eq:sup:equation for correlation function: intermediate}
\end{eqnarray}
where $\Delta_x$ is the Laplacian with respect to $\bm{x}$. Noting $C_{l}^{r,c()}(\bm{x},\bm{y};\dot{\gamma}) = C_{l}^{r,c()}(\bm{y},\bm{x};\dot{\gamma})$, we simplify Eq.~(\ref{eq:sup:equation for correlation function: intermediate}) as
\begin{eqnarray}
\Big\{\dot{\gamma} x_2\frac{\partial}{\partial x_1} + D_0 \Delta_x \Big(\Delta_x - r + c(x_2)\Big) \Big\} C^{r,c()}_{l}(\bm{x},\bm{y};\dot{\gamma}) = -D_0 T \Delta_x\delta(\bm{x}-\bm{y}) .
\label{eq:sup:equation for correlation function: c>0}
\end{eqnarray}
Now, we introduce the differential operators:
\begin{eqnarray}
\mathcal{L}_0(\bm{x},\bm{x}') &=& \Big[\dot{\gamma} x_2 \frac{\partial}{\partial x_1} + D_0 \Delta_x\Big(\Delta_x - r \Big)\Big] \delta(\bm{x}-\bm{x}') ,
\label{eq:sup:differential oparator1}
\end{eqnarray}
and rewrite Eq.~(\ref{eq:sup:equation for correlation function: c>0}) as
\begin{eqnarray}
\int d^3\bm{x}' \mathcal{L}_0(\bm{x},\bm{x}') C^{r,c()}_{l}(\bm{x}',\bm{y};\dot{\gamma}) = - D_0 \Delta_x c(x_2) C^{r,c()}_{l}(\bm{x},\bm{y};\dot{\gamma}) - D_0 T \Delta_x \delta(\bm{x}-\bm{y}) .
\label{eq:sup:equation for correlation function: c>0: rewrite}
\end{eqnarray}
Clearly, $\mathcal{L}_0(\bm{x},\bm{y})$ is connected with $C^{r,0}_{l}(\bm{x},\bm{y};\dot{\gamma})$ as
\begin{eqnarray}
\int d^3 \bm{x}'\mathcal{L}_0(\bm{x},\bm{x}')C^{r,0}_{l}(\bm{x}',\bm{y};\dot{\gamma}) = -D_0T \Delta_x \delta(\bm{x}-\bm{y}).
\label{eq:sup:equation for correlation function: c=0}
\end{eqnarray}
By interpreting Eq.~(\ref{eq:sup:equation for correlation function: c=0}) from a different viewpoint, we find that $C^{r,0}_{l}(\bm{x},\bm{y};\dot{\gamma})$ is identified with the inverse operator of $\mathcal{L}_0(\bm{x},\bm{y})$. Based on this observation, by acting the inverse operator $C^{r,0}_{l}(\bm{x},\bm{y};\dot{\gamma})$ on Eq.~(\ref{eq:sup:equation for correlation function: c>0: rewrite}), we obtain
\begin{eqnarray}
C^{r,c()}_{l}(\bm{x},\bm{y};\dot{\gamma}) = C^{r,0}_{l}(\bm{x},\bm{y};\dot{\gamma}) + \frac{1}{T} \int d^3 \bm{x}' C^{r,0}_{l}(\bm{x},\bm{x}';\dot{\gamma}) c(x'_2) C^{r,c()}_{l}(\bm{x}',\bm{y};\dot{\gamma}).
\label{eq:sup:act inverse EOM of ETCF: transient}
\end{eqnarray}
This equation connects $C^{r,0}_{l}(\bm{x},\bm{y};\dot{\gamma})$ with $C^{r,c()}_{l}(\bm{x},\bm{y};\dot{\gamma})$.

When $c(x_2)$ and $C^{r,c()}_{l}(\bm{x},\bm{y};\dot{\gamma})$ have simple forms, we can derive more convenient relations for $C^{r,c()}_{l}(\bm{x},\bm{y};\dot{\gamma})$ from Eq.~(\ref{eq:sup:act inverse EOM of ETCF: transient}). The simplest example is $c(x_2)=c_0 \delta(x_2)$, for which we can immediately derive the simpler equation for $C^{r,c()}_{l}(\bm{k}_{\parallel},x_2=0,y_2=0,\dot{\gamma})$ as
\begin{eqnarray}
C^{r,c()}_{l}(\bm{k}_{\parallel},0,0;\dot{\gamma}) = \frac{TC^{r,0}_{l}(\bm{k}_{\parallel},0,0;\dot{\gamma})}{T-c_0C^{r,0}_{l}(\bm{k}_{\parallel},0,0;\dot{\gamma})}.
\label{eq:sup:expression of correlation function of x2=0: delta}
\end{eqnarray}

As another example, we consider that $c(x_2)$ and $C^{r,c()}_{l}(\bm{k}_{\parallel},x_2,0;\dot{\gamma})$ have exponential forms:
\begin{eqnarray}
c(x_2) &=& c_0 \delta(x_2) + \sum_{i=1}^n c^{\rm (i)}_1e^{-x_2/\ell^{\rm (i)}_1}, 
\label{eq:sup:exponential ansatz of c}\\
C^{r,c()}_{l}(\bm{k}_{\parallel},x_2,0;\dot{\gamma}) &=& C^{r,c()}_{l}(\bm{k}_{\parallel},0,0;\dot{\gamma}) e^{-x_2/\ell_2[c()]}.
\label{eq:sup:exponential ansatz of correlation function}
\end{eqnarray}
By substituting Eqs.~(\ref{eq:sup:exponential ansatz of c}) and (\ref{eq:sup:exponential ansatz of correlation function}) into Eq.~(\ref{eq:sup:act inverse EOM of ETCF: transient}), we obtain
\begin{eqnarray}
C^{r,c()}_{l}(\bm{k}_{\parallel},0,0;\dot{\gamma}) = \frac{TC^{r,0}_{l}(\bm{k}_{\parallel},0,0;\dot{\gamma})}{T-\bar{c}C^{r,0}_{l}(\bm{k}_{\parallel},0,0;\dot{\gamma})}
\label{eq:sup:expression of correlation function of x2=0: exponential}
\end{eqnarray}
with 
\begin{eqnarray}
\bar{c} &=& c_0 + \sum_{i=1}^N c_1^{\rm (i)}\ell^{\rm (i)}_{\rm sum} ,\nonumber \\[3pt]
\ell^{\rm (i)}_{\rm sum} &=& \ell_1^{\rm (i)}+\ell_2(0)+\ell_2[c()].
\end{eqnarray}
The case $n=1$ corresponds to Eq.~(15) in the main text.

\subsection{Derivation of Eq.~(27)}
We here study the case $c(x_2)=0$, which corresponds to the dynamics in the bulk region. The starting point of our analysis is an exact integral expression for $C^{r,0}_{l}(\bm{k};\dot{\gamma})$:
\begin{eqnarray}
C^{r,0}_{l}(\bm{k};\dot{\gamma}) = TD_0 \int_0^{\infty} ds e^{-D_0\int_0^s d\lambda |\bm{\kappa}_{\lambda}|^{2}( |\bm{\kappa}_{\lambda}|^{2} + r)} |\bm{\kappa}_s|^{2}
\label{eq:sup:formal expression of correlation function in the bulk: appendix}
\end{eqnarray}
with $\bm{\kappa}_\lambda = (k_1,k_2+\dot{\gamma} \lambda k_1/2,k_3)$. This type of expression was initially derived by Onuki and Kawasaki~\cite{sup:onuki1979nonequilibrium}, and has been widely used in the analysis of fluctuations in the presence of shear flow. We recently summarized its compact derivation in Ref.~\cite{sup:nakano2021long}. Therefore, we omit the details of derivation and study the asymptotic expression in the long-wavelength region.

First, by expanding the $\lambda$-integral of Eq.~(\ref{eq:sup:formal expression of correlation function in the bulk: appendix}), we obtain
\begin{eqnarray}
C^{r,0}_{l}(\bm{k};\dot{\gamma}) = D_0 T \int_0^{\infty} ds \Big[k_1^2 + (k_2+\frac{1}{2} \dot{\gamma} s k_1)^2 + k_3^2 \Big] e^{-F_B(s;\bm{k})}
\label{eq:sup:correlation function in bulk for conserved non-equilibrium: explicit form}
\end{eqnarray}
with
\begin{eqnarray}
F_B(s;\bm{k}) &=& D_0 \Big[s r \Big( |\bm{k}|^2 + \frac{1}{2} s \dot{\gamma} k_1 k_2 + \frac{1}{12} s^2 \dot{\gamma}^2 k_1^2 \Big) \nonumber \\[3pt]
&+& s \Big(|\bm{k}|^4 + s k_1 k_2 (\dot{\gamma} |\bm{k}|^2 + \frac{1}{8} \dot{\gamma}^3 s^2 k_1^2) + \frac{1}{6} \dot{\gamma}^2 s^2 k_1^2 (|\bm{k}|^2 + 2 k_2^2) + \frac{1}{80} s^4 \dot{\gamma}^4 k_1^4\Big)\Big].
\label{eq:sup:GammaB integral}
\end{eqnarray}
Noting
\begin{eqnarray}
\frac{\partial}{\partial s} e^{-F_B(s;\bm{k})} = -D_0 |\bm{\kappa}_{s}|^{2}( |\bm{\kappa}_{s}|^{2} + r)e^{-F_B(s;\bm{k})},
\end{eqnarray}
we have
\begin{eqnarray}
C_l^{r,0}(\bm{k},\dot{\gamma}) = - T \int_0^{\infty} ds \frac{1}{|\bm{\kappa}_{s}|^{2}+r} \frac{\partial}{\partial s} e^{-F_B(s;\bm{k})}.
\label{eq:sup:correlation function in bulk for conserved non-equilibrium: transit1}
\end{eqnarray}
Using integration by parts, Eq.~(\ref{eq:sup:correlation function in bulk for conserved non-equilibrium: transit1}) is rewritten as
\begin{eqnarray}
C^{r,0}_{l}(\bm{k};\dot{\gamma}) = \frac{T}{r + |\bm{k}|^2} - \dot{\gamma} T \int_0^{\infty} ds e^{-F_B(s;\bm{k})} \frac{ k_1(k_2 + \frac{1}{2} s \dot{\gamma} k_1) }{\Big(r + |\bm{k}|^2 + s \dot{\gamma} k_1 k_2 + \frac{1}{4} s^2 \dot{\gamma}^2 k_1^2\Big)^{2}} .
\label{eq:sup:correlation function in bulk for conserved non-equilibrium: another form}
\end{eqnarray}

For $|\bm{k}| \ll r^{1/2}$, we can neglect the terms of order $|\bm{k}|^4$ in Eq.~(\ref{eq:sup:correlation function in bulk for conserved non-equilibrium: another form}) and obtain the approximation form
\begin{eqnarray}
C^{r,0}_{l}(\bm{k};\dot{\gamma}) \simeq \frac{T}{r + |\bm{k}|^2} - \frac{\dot{\gamma} T}{r^2}\int_0^{\infty} ds\Big(k_1 k_2 + \frac{1}{2} s \dot{\gamma} k_1^2 \Big) e^{-D_0  s r \Big( |\bm{k}|^2 + \frac{1}{2} s \dot{\gamma} k_1 k_2 + \frac{1}{12} s^2 \dot{\gamma}^2 k_1^2 \Big)} .
\label{eq:sup:correlation function in bulk for conserved non-equilibrium: approximated form}
\end{eqnarray}
To obtain the asymptotic behavior of $C^{r,0}_{l}(\bm{k};\dot{\gamma})$ from this expression, we divide the $\bm{k}$-region into two regions
\begin{eqnarray}
{\rm (i)} \ \  \frac{1}{12} \dot{\gamma}^2 k_1^2 \ll D_0^3 r^3 |\bm{k}|^6, \nonumber \\[3pt]
{\rm (ii)} \ \   \frac{1}{12} \dot{\gamma}^2 k_1^2 \gg D_0^3 r^3 |\bm{k}|^6.
\end{eqnarray}
In region (i), the dominant contribution of the $s$-integral arises from
\begin{eqnarray}
s = \frac{1}{D_0 r |\bm{k}|^2},
\end{eqnarray}
where the integrand is approximated as
\begin{eqnarray}
\Big(k_1 k_2 + \frac{1}{2} s \dot{\gamma} k_1^2 \Big) e^{-D_0  s r \Big( |\bm{k}|^2 + \frac{1}{2} s \dot{\gamma} k_1 k_2 + \frac{1}{12} s^2 \dot{\gamma}^2 k_1^2 \Big)}  \simeq \Big(k_1 k_2 + \frac{1}{2} s \dot{\gamma} k_1^2 \Big) e^{- D_0  s r |\bm{k}|^2}.
\end{eqnarray}
Then, Eq.~(\ref{eq:sup:correlation function in bulk for conserved non-equilibrium: approximated form}) is approximately rewritten as
\begin{eqnarray}
C^{r,0}_{l}(\bm{k};\dot{\gamma}) \simeq \frac{T}{r} - \frac{T}{r^2} |\bm{k}|^2 - \frac{\dot{\gamma}T}{D_0 r^3} \frac{k_1 k_2}{|\bm{k}|^2} - \frac{\dot{\gamma}^2T}{2D_0^2 r^4} \frac{k_1^2}{ |\bm{k}|^4} -\cdots.
\label{eq:sup:asmptotic behavior in region i: model B bulk}
\end{eqnarray}
In region (ii), the dominant contribution of the $s$-integral arises from
\begin{eqnarray}
s = \Big(\frac{12}{D_0 \dot{\gamma} r k_1^2}\Big)^{\frac{1}{3}},
\end{eqnarray}
where the integrand is approximated as
\begin{eqnarray}
\Big(k_1 k_2 + \frac{1}{2} s \dot{\gamma} k_1^2 \Big) e^{-D_0  s r \Big( |\bm{k}|^2 + \frac{1}{2} s \dot{\gamma} k_1 k_2 + \frac{1}{12} s^2 \dot{\gamma}^2 k_1^2 \Big)}  \simeq \Big(k_1 k_2 + \frac{1}{2} s \dot{\gamma} k_1^2 \Big) e^{-\frac{1}{12}  D_0  r s^3 \dot{\gamma}^2 k_1^2}.
\end{eqnarray}
Then, Eq.~(\ref{eq:sup:correlation function in bulk for conserved non-equilibrium: approximated form}) is approximately rewritten as
\begin{eqnarray}
C^{r,0}_{l}(\bm{k};\dot{\gamma}) \simeq \frac{T}{r} - \frac{12^{\frac{2}{3}}}{6} \Gamma\Big(\frac{2}{3}\Big) T \frac{\dot{\gamma}^{\frac{2}{3}}}{r^{\frac{8}{3}}D_0^{\frac{2}{3}}} |k_1|^{\frac{2}{3}} - 12^{\frac{1}{3}} \Gamma\Big(\frac{4}{3}\Big) T \frac{\dot{\gamma}^{\frac{1}{3}}}{r^{\frac{7}{3}}D_0^{\frac{1}{3}}} |k_1|^{\frac{1}{3}}k_2 \cdots .
\label{eq:sup:asmptotic behavior in region ii: model B bulk}
\end{eqnarray}

The behaviors of Eqs.~(\ref{eq:sup:asmptotic behavior in region i: model B bulk}) and (\ref{eq:sup:asmptotic behavior in region ii: model B bulk}) are understood as two limiting cases of the following equation
\begin{eqnarray}
C^{r,0}_{l}(\bm{k};\dot{\gamma}) = \frac{T}{r + |\bm{k}|^2 + c_B \{(\dot{\gamma}/r D_0)|k_1|\}^{2/3}+\cdots},
\label{eq:sup:correlation function: c0=0: conserved}
\end{eqnarray}
where $c_B\simeq1.18$. If we use the typical length scales $l_B=(D_0/\dot{\gamma})^{1/4}$ and $\xi_B=\sqrt{1/r}$, Eq.~(\ref{eq:sup:correlation function: c0=0: conserved}) is rewritten as
\begin{eqnarray}
C^{r,0}_{l}(\bm{k};\dot{\gamma}) = \frac{T}{r + |\bm{k}|^2 + c_B \{(\xi^2_B/l^4_B)|k_1|\}^{2/3}+\cdots}.
\end{eqnarray}
Here, $l_B$ is the characteristic length of the flow, and $\xi_B$ is the correlation length of fluctuations in the equilibrium system. 

To check the validity of the above calculation, we numerically integrate Eq.~(\ref{eq:sup:correlation function in bulk for conserved non-equilibrium: another form}). The result is presented in Fig.~\ref{fig: bulk correlation function: B sr=0.3 r=10}. The parameter settings are the same as in Fig.~3. We find the good agreement between Eq.~(\ref{eq:sup:correlation function: c0=0: conserved}) and the numerical result.
\begin{figure}[h]
\begin{center}
\includegraphics[width=8cm]{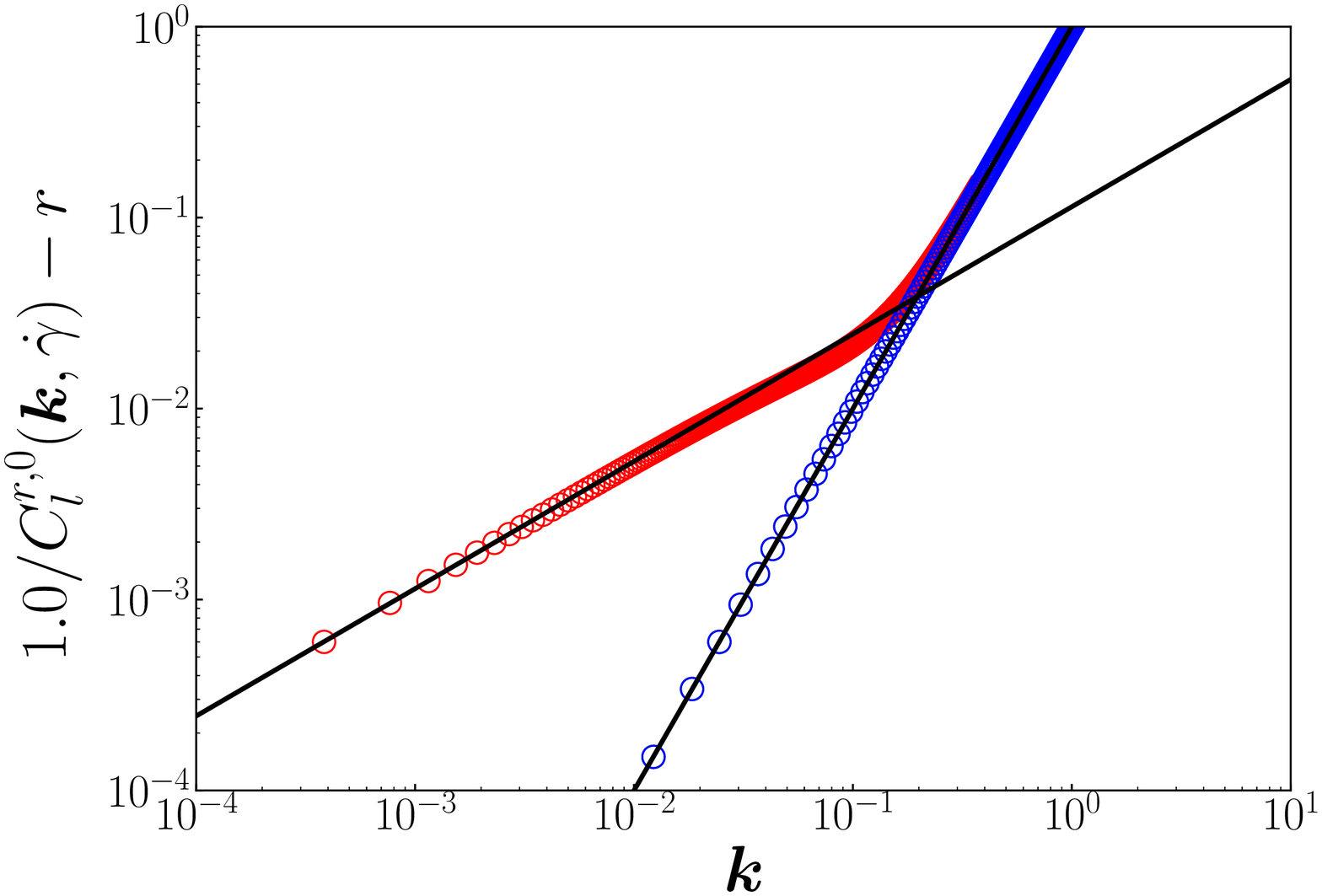}
\end{center}
\vspace{-0.5cm}
\caption{Red: $1.0/C^{r,0}_l(k_1,k_2=k_3=0;\dot{\gamma})-r$ vs. $k_1$. Blue: $1.0/C^{r,0}_l(k_1=0,k_2,k_3=0;\dot{\gamma})-r$ vs. $k_2$. The black lines are, respectively, given by $1.0/C^{r,0}_l(k_1,k_2=k_3=0;\dot{\gamma})-r = c_B \{(\xi^2_B/l^4_B)|k_1|\}^{2/3}$, and $1.0/C^{r,0}_l(k_1=0,k_2,k_3=0;\dot{\gamma})-r = k_2^2$.}
\label{fig: bulk correlation function: B sr=0.3 r=10}
\end{figure}

\subsection{Derivation of Eq.~(24)}
As previously shown, when $c(x_2)$ and $C^{r,c()}_{l}(\bm{x},\bm{y};\dot{\gamma})$ satisfy Eqs.~(\ref{eq:sup:exponential ansatz of c}) and (\ref{eq:sup:exponential ansatz of correlation function}), $C^{r,c()}_{l}(\bm{k}_{\parallel},0,0;\dot{\gamma})$ is written as Eq.~(\ref{eq:sup:expression of correlation function of x2=0: exponential}). Here, we derive the asymptotic expression of $C^{r,c()}_{l}(\bm{k}_{\parallel},0,0;\dot{\gamma})$ from Eq.~(\ref{eq:sup:expression of correlation function of x2=0: exponential}). Because Eq.~(\ref{eq:sup:expression of correlation function of x2=0: exponential}) connects $C^{r,c()}_{l}(\bm{k}_{\parallel},0,0;\dot{\gamma})$ with $C^{r,0}_{l}(\bm{k}_{\parallel},0,0;\dot{\gamma})$, we first calculate the asymptotic form of $C^{r,0}_{l}(\bm{k}_{\parallel},0,0;\gamma)$ for $|\bm{k}_\parallel|\ll r^{1/2}$. By taking the Fourier transform of Eq.~(\ref{eq:sup:correlation function in bulk for conserved non-equilibrium: approximated form}) with respect to the $x_2$-coordinate and substituting $x_2=0$, we obtain
\begin{eqnarray}
C^{r,0}_{l}(\bm{k}_{\parallel},0,0;\dot{\gamma}) \simeq \int_{-\infty}^{\infty}\frac{dk_2}{2\pi}\frac{T}{r + \bm{k}^2} - \frac{\dot{\gamma} T}{r^2} \int_{-\infty}^{\infty}\frac{dk_2}{2\pi}\int_0^{\infty}ds \Big(k_1k_2+\frac{1}{2}\dot{\gamma} s k_1^2\Big) e^{-D_0 r\big(s|\bm{k}|^2+\frac{1}{2}\dot{\gamma} s^2 k_1 k_2 + \frac{1}{12}\dot{\gamma}^2 s^3 k_1^2 \big)} .\nonumber \\
\label{eq:sup:correlation function: under shear c=0 model b:mod1}
\end{eqnarray}
Noting that Eq.~(\ref{eq:sup:correlation function in bulk for conserved non-equilibrium: approximated form}) holds for $|\bm{k}|\ll r^{1/2}$, we find that this expression is valid for $|\bm{k}_{\parallel}| \ll r^{1/2}$. The $k_2$-integral in Eq.~(\ref{eq:sup:correlation function: under shear c=0 model b:mod1}) is explicitly calculated as
\begin{eqnarray}
C^{r,0}_{l}(\bm{k}_{\parallel},0,0;\dot{\gamma}) &\simeq& \frac{T}{2}\sqrt{\frac{1}{r + \bm{k}_{\parallel}^2}} - \frac{T}{8\sqrt{\pi}} \frac{\dot{\gamma}^2 k_1^2}{\sqrt{D_0 r^5}} \int_0^{\infty}ds s^{\frac{1}{2}} e^{-D_0 r \big(s\bm{k}_{\parallel}^2 + \frac{1}{48}\dot{\gamma}^2 s^3 k_1^2 \big)} .
\label{eq:sup:correlation function: under shear c=0 model b:mod2}
\end{eqnarray}
The $s$-integral is calculated in the similar way as in bulk. We omit the details; the result is given by
\begin{eqnarray}
C^{r,0}_{l}(\bm{k}_{\parallel},0,0;\dot{\gamma}) =
\begin{cases}
\frac{T}{2\sqrt{r}} - \frac{T}{4r\sqrt{r}} |\bm{k}_{\parallel}|^2  - \frac{\dot{\gamma}T}{2\sqrt{3} D_0 r^3}|k_1| + \cdots \ \ {\rm for} \ 4\sqrt{3}D_0r |\bm{k}_{\parallel}|^3 \ll \dot{\gamma} |k_1|, \\
\frac{T}{2\sqrt{r}} - \frac{T}{4r\sqrt{r}} |\bm{k}_{\parallel}|^2 - \frac{T\dot{\gamma}^2}{16 D_0^2 r^4} \frac{k_1^2}{|\bm{k}_{\parallel}|^3} \cdots \ \ {\rm for} \ 4\sqrt{3}D_0r |\bm{k}_{\parallel}|^3 \gg \dot{\gamma} |k_1|.
 \end{cases}
 \label{eq:sup:correlation function: under shear c=0 model b:final}
\end{eqnarray}
Then, by substituting Eq.~(\ref{eq:sup:correlation function: under shear c=0 model b:final}) into Eq.~(\ref{eq:sup:expression of correlation function of x2=0: exponential}), we obtain 
\begin{eqnarray}
C^{r,c()}_{l}(\bm{k}_{\parallel},0,0;\dot{\gamma}) = \frac{T}{(2\sqrt{r}-\bar{c}) + \frac{2}{\sqrt{3}}\frac{\dot{\gamma}}{r^2 D_0} |k_1| + \frac{1}{\sqrt{r}} |\bm{k}_{\parallel}|^2 + \cdots},
\end{eqnarray}
where we have ignored higher order terms than $O(\dot{\gamma})$. Finally, by taking as $\bar{c}=2\sqrt{r}$, we obtain Eq.~(24).

\section{Supplemental Numerical Analysis}

\subsection{Renormalization of $r$}
We consider the renormalization of $r$. As explained in the main text, it is calculated by self-consistently solving the following equation
\begin{eqnarray}
r = r_0 + g \int_{2\pi/L}^{2\pi/a^{\rm uv}} \frac{d^3\bm{k}}{(2\pi)^3} C_l^{r,0}(\bm{k};\dot{\gamma}),
\label{eq:sup:self-consistent equation of r: supple}
\end{eqnarray}
where $a^{\rm uv} = a^{\rm uv}_1 = a^{\rm uv}_2 = a^{\rm uv}_3$.
We numerically solve this equation and obtain $r$ as a function of $r_0$. The result is presented in Fig.~\ref{fig:renormalization of r}. The parameter values are chosen as $T=g=1.0$, $L_1=L_2=L_3=512.0$ and $a^{\rm uv} =1.0$.
\begin{figure}[h]
\begin{center}
\includegraphics[width=8cm]{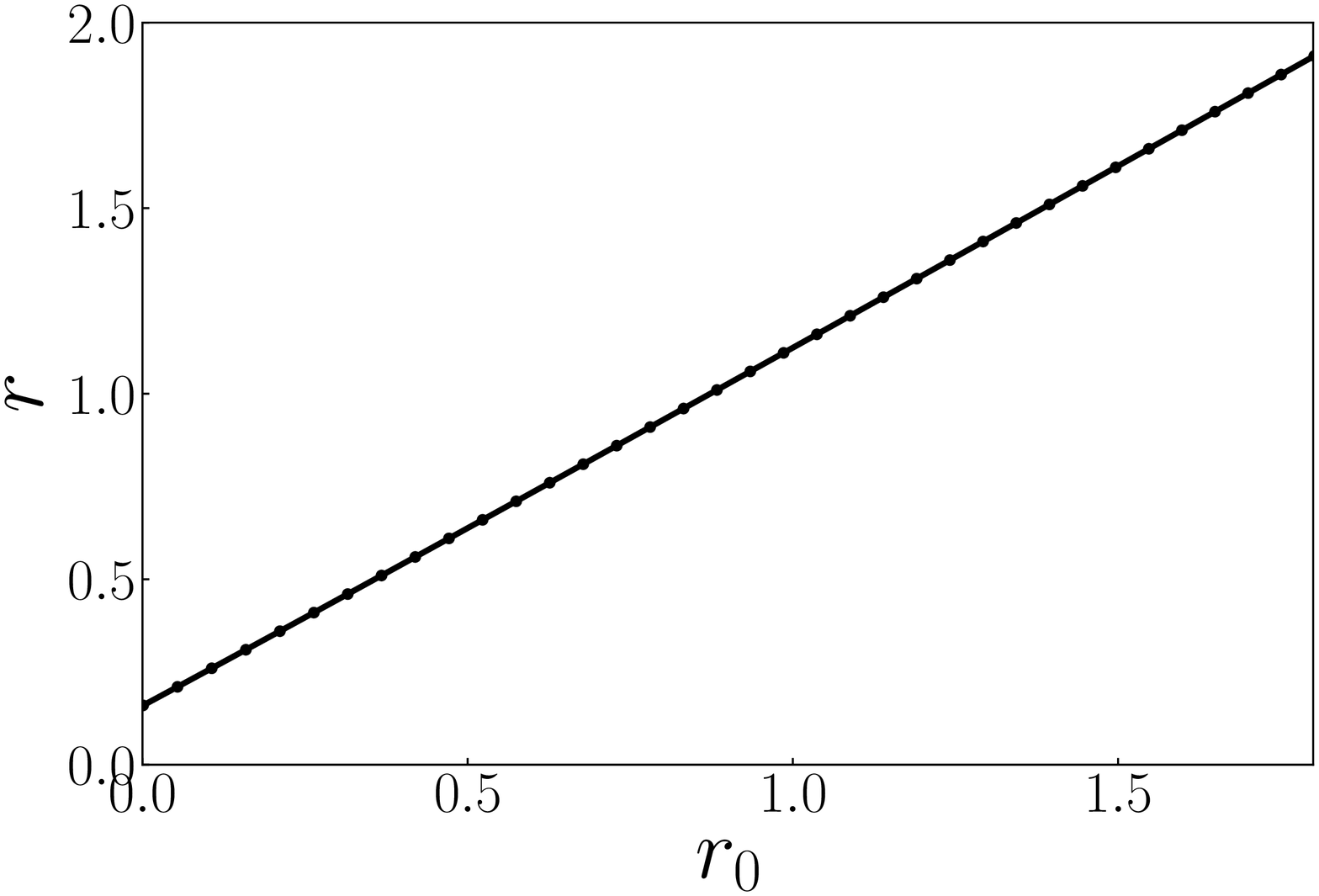}
\end{center}
\vspace{-0.5cm}
\caption{$r$ as a function of $r_0$, which is obtained by numerically solving Eq.~(\ref{eq:sup:self-consistent equation of r: supple}).}
\label{fig:renormalization of r}
\end{figure}
We note that the value of $r$ depends on the choice of $a^{\rm uv}$. Generally, $r$ diverges as $a^{\rm uv}$ approaches $+0$. As shown in Fig.~\ref{fig:renormalization of r}, the renormalization effect of $r$ is rather small when $a^{\rm uv}$ is set to $1.0$.

\subsection{How to draw the guideline in Fig.~4}
In the main text, we derived the analytical expression Eq.~(26) for the critical point $c_0^{sc}(r_0;\dot{\gamma},T)$. This expression is valid for $\dot{\gamma} \to +0$ and $a^{\rm uv}_2 \to 0$. However, we drew the phase diagram Fig.~4 with using $a^{\rm uv}_2=1.0$ to reduce the numerical cost. Here, we argue the influence of the finite cutoff.

First of all, we numerically calculate $C^r_{\rm sc}(\bm{k}_{\parallel};\dot{\gamma})$ with $a^{\rm uv}_2 = 1.0$ using Eq.~(19) and (23), and plot it as the red and blue plots in Fig.~\ref{fig: plane correlation function at criticality using auv=1.0}. The red and blue plots, respectively, give $C^r_{\rm sc}(k_1,k_3=0;\dot{\gamma})$ as a function of $k_1$ and $C^r_{\rm sc}(k_1=0,k_3;\dot{\gamma})$ as a function of $k_3$. The black solid curve is obtained by fitting with
\begin{eqnarray}
C^{r}_{\rm sc}(\bm{k}_{\parallel};\dot{\gamma}) = A \frac{T}{(2 \dot{\gamma}/\sqrt{3}D_0 r^{2})|k_1| + |\bm{k}_{\parallel}|^2/\sqrt{r}},
\end{eqnarray}
where $A$ is the fitting parameter. By fitting the numerical data with $|\bm{k}|<0.1$, we obtain $A=3.118$. We find that the red and blue plots agree well with the black solid curves except for a slightly difference in the long-wavelength region. We note that there is no such a deviation for $a^{\rm uv}_2=0.01$ (see Fig.~3 in the main text).
\begin{figure}[h]
\begin{center}
\includegraphics[width=8cm]{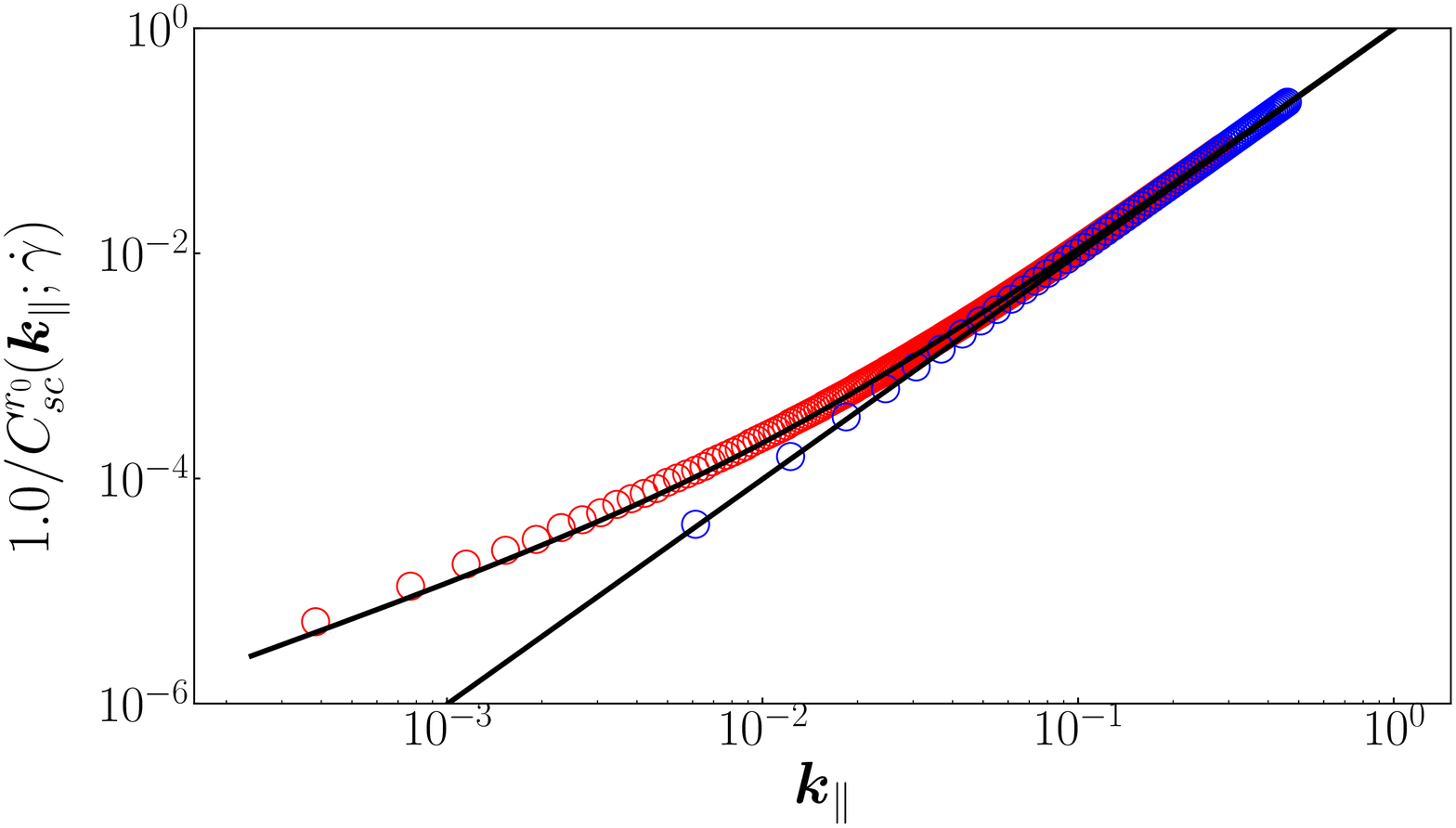}
\end{center}
\vspace{-0.5cm}
\caption{Same as Fig.~3 in the main text, but with $a_{\rm uv}= 1.0$.}
\label{fig: plane correlation function at criticality using auv=1.0}
\end{figure}

By considering the additional factor $A$, the expression Eq.~(26) is modified to
\begin{eqnarray}
c_0^{\rm sc}(r_0;\dot{\gamma},T) &\simeq& c_{\rm max} + Ag l_{\rm sum} \int_{2\pi/L}^{2\pi/a^{\rm uv}} \frac{d^2\bm{k}_{\parallel}}{(2\pi)^2}\frac{T}{(2 \dot{\gamma}/\sqrt{3}D_0 r^{2})|k_1| + |\bm{k}_{\parallel}|^2/\sqrt{r}} - \ell_{\rm sum} (r-r_0)\nonumber \\[3pt]
&\simeq& c_{\rm max} + A g l_{\rm sum} \int_{k_c}^{2\pi/a^{\rm uv}} \frac{d^{2}\bm{k}_{\parallel}}{(2\pi)^{2}} \frac{T}{|\bm{k}_{\parallel}|^2/\sqrt{r}} - \ell_{\rm sum} (r-r_0) \nonumber \\[3pt]
&=& c_{\rm max} - \frac{g l_{\rm sum}T A \sqrt{r}}{2\pi} \log \big(\frac{\Lambda \dot{\gamma}}{\sqrt{3}\pi D_0r_0^{3/2}}\big) - \ell_{sum} (r-r_0).
\label{eq:sup:transition point: explicit form under shear: plus A factor}
\end{eqnarray}
The guideline in Fig.~4 is drawn by using Eq.~(\ref{eq:sup:transition point: explicit form under shear: plus A factor}). Concretely, the functional form of the guideline is given by
\begin{eqnarray}
c_0 - c_{\rm max}  = 0.74 - 1.569 \log \dot{\gamma},
\end{eqnarray}
where we have used the parameter given in Fig.~4 and $0.74$ is adjusted by eyes. Figure~4 shows that its slope agrees well with Eq.~(\ref{eq:sup:transition point: explicit form under shear: plus A factor}) as expected.

\section{Non-conserved case}
In the main text, we studied the model where the order parameter is conserved in the time evolution. As a related model, we can also consider the dynamics where the order parameter is not conserved. It is given by the following equation:
\begin{eqnarray}
\frac{\partial \varphi^a(\bm{x},t)}{\partial t} +  \bm{v}(\bm{x}) \cdot \nabla \varphi^a(\bm{x},t) = - \Gamma_0 \frac{\delta \Phi[\bm{\varphi}]}{\delta \varphi^a(\bm{x},t)} + \eta^a(\bm{x},t),
\label{eq:sup:time dependent Ginzburg-Landau model: non-conserved}
\end{eqnarray}
where $\Gamma_0$ is a bare diffusion constant and $\bm{\eta}(\bm{x},t)$ is the Gaussian white noise satisfying
\begin{eqnarray}
\big\langle \eta^a(\bm{x},t) \big\rangle &=& 0, \\
\big\langle \eta^a(\bm{x},t)\eta^b(\bm{x}',t') \big\rangle &=& 2T\delta_{a b} \Gamma_0 \delta(\bm{x}-\bm{x}') \delta(t-t').
\label{eq:sup:bare transport coefficient: explicit form: non-conserved}
\end{eqnarray}
The Landau--Ginzburg free energy $\Phi[\bm{\varphi}]$ is the same as that of the conserved case. We note that in equilibrium, the non-conserved and conserved dynamics correspond to model A and B in the classification of Hohenberg and Halperin~\cite{sup:hohenberg1977theory}, respectively.

In contrast to the conserved case, there is no localized long-range order for the non-conserved case. Here, we derive this result.

\subsection{Fluctuations in the disordered bulk}
First, we study linear fluctuations in the disordered bulk. For the non-conserved case, $C^{r,0}_{l}(\bm{k};\dot{\gamma})$ is given by
\begin{eqnarray}
C^{r,0}_{l}(\bm{k};\dot{\gamma}) =  \Gamma_0 T \int_0^{\infty}ds e^{-\Gamma_0 \big\{s(r + |\bm{k}|^2)+\frac{1}{2}\dot{\gamma} s^2 k_1 k_2 + \frac{1}{12}\dot{\gamma}^2 s^3 k_1^2 \big\}}.
\label{eq:sup:correlation function in bulk for non-conserved non-equilibrium: explicit form}
\end{eqnarray}
When we restrict ourselves to the long-wavelength region $|\bm{k}| \ll r^{1/2}$, the dominant contribution of the $s$-integral comes from $s\sim 1/\Gamma_0r$. Accordingly, we can simply expand Eq.~(\ref{eq:sup:correlation function in bulk for non-conserved non-equilibrium: explicit form}) as
\begin{eqnarray}
C^{r,0}_{l}(\bm{k};\dot{\gamma}) &=& \Gamma_0 T \int_0^{\infty}ds  \Big(1 -\Gamma_0 s|\bm{k}|^2-\frac{\Gamma_0}{2}\dot{\gamma} s^2 k_1 k_2  -\frac{\Gamma_0}{12} \dot{\gamma}^2 s^3 k_1^2 \cdots \Big) e^{-\Gamma_0 s r} \nonumber \\
&=& \frac{T}{r} - \frac{T}{r^2} |\bm{k}|^2 - \frac{T\dot{\gamma}}{\Gamma_0r^3} k_1k_2 - \frac{1}{2}\frac{T\dot{\gamma}^2}{\Gamma_0^2r^4}k_1^2\ \cdots.
\end{eqnarray}
To make it easier to see, we rewrite it as
\begin{eqnarray}
C^{r,0}_{l}(\bm{k};\dot{\gamma}) \simeq \frac{T}{r + |\bm{k}|^2 + (\dot{\gamma}/r\Gamma_0) k_1k_2 +  (\dot{\gamma}^2/2r^2\Gamma_0^2)k_1^2}.
\label{eq:sup:correlation function: c0=0: non-conserved}
\end{eqnarray}
If we use the typical length scales $l_A=\sqrt{\Gamma_0/\dot{\gamma}}$ and $\xi_A=\sqrt{1/r}$, Eq.~(\ref{eq:sup:correlation function: c0=0: non-conserved}) is rewritten as
\begin{eqnarray}
C^{r,0}_{l}(\bm{k};\dot{\gamma}) \simeq \frac{T}{r + |\bm{k}|^2 + (\xi_A/l_A)^2 k_1k_2 + (\xi_A/l_A)^4 k^2_1/2}.
\label{eq:sup:correlation function in the disordered bulk: non-conserved case}
\end{eqnarray}
Eq.~(\ref{eq:sup:correlation function in the disordered bulk: non-conserved case}) corresponds to Eq.~(27) for the conserved case. We also present the result obtained by numerically integrating Eq.~(\ref{eq:sup:correlation function in bulk for non-conserved non-equilibrium: explicit form}) in the left-hand side of Fig.~\ref{fig: correlation function: A sr=1 r=1}. This figure supports the validity of Eq.~(\ref{eq:sup:correlation function: c0=0: non-conserved}).

Equation~(\ref{eq:sup:correlation function in the disordered bulk: non-conserved case}) implies that the shear flow does not lead to the anomalous suppression although it makes the fluctuation anisotropic. Accordingly, because there is no effective force to stabilize the long-range order in contrast to the conserved case, it is predicted that the two-dimensional localized long-range order does not occur.

\subsection{Proof of non-existence of surface long-range order}
In the same way as the conserved case, the expression of the transition point is given by Eq.~(22). Below, we show the non-existence of surface long-range order by demonstrating that the infrared divergence is not removed for the non-conserved case.

We calculate the asymptotic expression of $C_{sc}^r(\bm{k}_{\parallel};\dot{\gamma})$ for $|\bm{k}_{\parallel}| \ll r^{1/2}$ from Eq.~(\ref{eq:sup:correlation function in bulk for non-conserved non-equilibrium: explicit form}). By taking the Fourier transform of Eq.~(\ref{eq:sup:correlation function in bulk for non-conserved non-equilibrium: explicit form}) with respect to the $x_2$-coordinate and substituting $x_2=0$, we have
\begin{eqnarray}
C^{r,0}_{l}(\bm{k}_{\parallel},0,0;\dot{\gamma}) =  \frac{T}{2}\sqrt{\frac{\Gamma_0}{\pi}} \int_0^{\infty}ds~s^{-\frac{1}{2}} e^{-\Gamma_0 \big\{s(r+|\bm{k}_{\parallel}|^2)+ \frac{1}{48}\dot{\gamma}^2 s^3 k_1^2 \big\}}. 
\label{eq:sup:appendix c2: int}
\end{eqnarray}
For $|\bm{k}_{\parallel}| \ll r^{1/2}$, the integrand of (\ref{eq:sup:appendix c2: int}) is expanded as
\begin{eqnarray}
C^{r,0}_{l}(\bm{k}_{\parallel},0,0;\dot{\gamma}) =  \frac{T}{2}\sqrt{\frac{\Gamma_0}{\pi}} \int_0^{\infty}ds~s^{-\frac{1}{2}} e^{-\Gamma_0 r s}  \Big(1- \Gamma_0 s |\bm{k}_{\parallel}|^2 - \frac{1}{48} \Gamma_0 \dot{\gamma}^2 s^3 k_1^2 + \cdots \Big). 
\label{eq:sup:appendix c2: intint}
\end{eqnarray}
Integrating each term of (\ref{eq:sup:appendix c2: intint}) with respect to $s$ yields
\begin{eqnarray}
C^{r,0}_{l}(\bm{k}_{\parallel},0,0;\dot{\gamma}) = \frac{T}{2\sqrt{r_0}} \Big(1- \frac{1}{2r} |\bm{k}_{\parallel}|^2 - \frac{15}{384}\frac{\dot{\gamma}^2}{\Gamma_0^2 r^3} k_1^2 + \cdots \Big).
\label{eq:sup:correlation function: under shear c=0 model a:mod1}
\end{eqnarray}
Finally, by noting that Eq.~(15) also holds for the non-conserved case, we obtain
\begin{eqnarray}
C_{\rm sc}^r(\bm{k}_{\parallel};\dot{\gamma}) = \frac{T}{|\bm{k}_{\parallel}|^2/\sqrt{r} + \{15\dot{\gamma}^2/(192\Gamma_0^2 r^{5/2})\} k_1^2 + \cdots } .
\label{eq:sup:correlation function: under shear c=0 model a: final}
\end{eqnarray}
This result is checked by comparing with the one numerically integrating Eq.~(\ref{eq:sup:appendix c2: int}). It is presented in the right-hand side of Fig.~\ref{fig: correlation function: A sr=1 r=1}. We find the good agreement between both results.

This expression implies that the shear flow stretches the fluctuation in the $x_2=0$ plane along the $x_1$-axis, but does not lead to the anomalous suppression. Therefore, the shear flow does not remove the infrared divergence, which is the same as in the equilibrium case. Thus, we conclude that the two-dimensional localized long-range order does not appear for the non-conserved case.

\begin{figure}[h]
\begin{center}
\includegraphics[width=8cm]{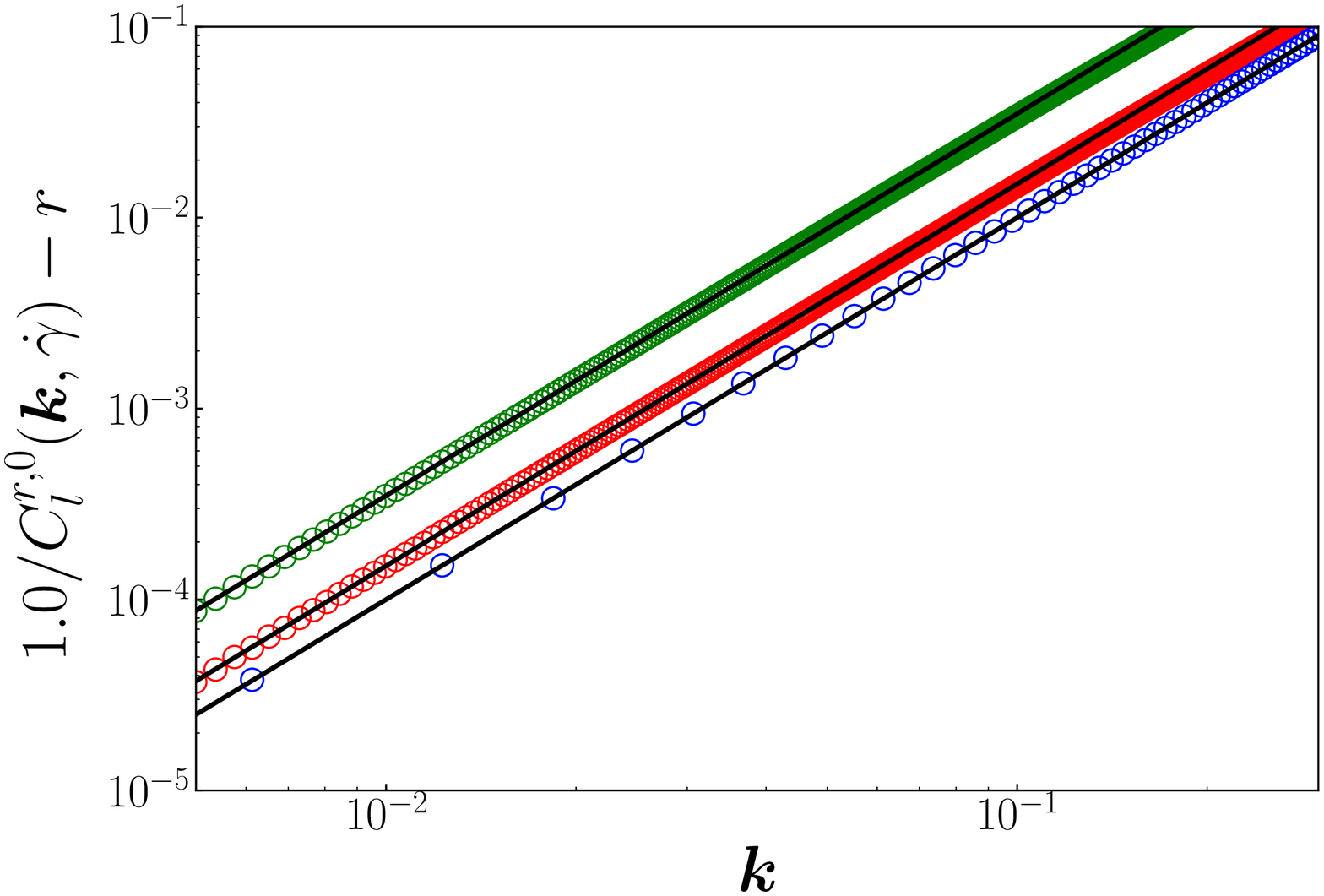}
\includegraphics[width=8cm]{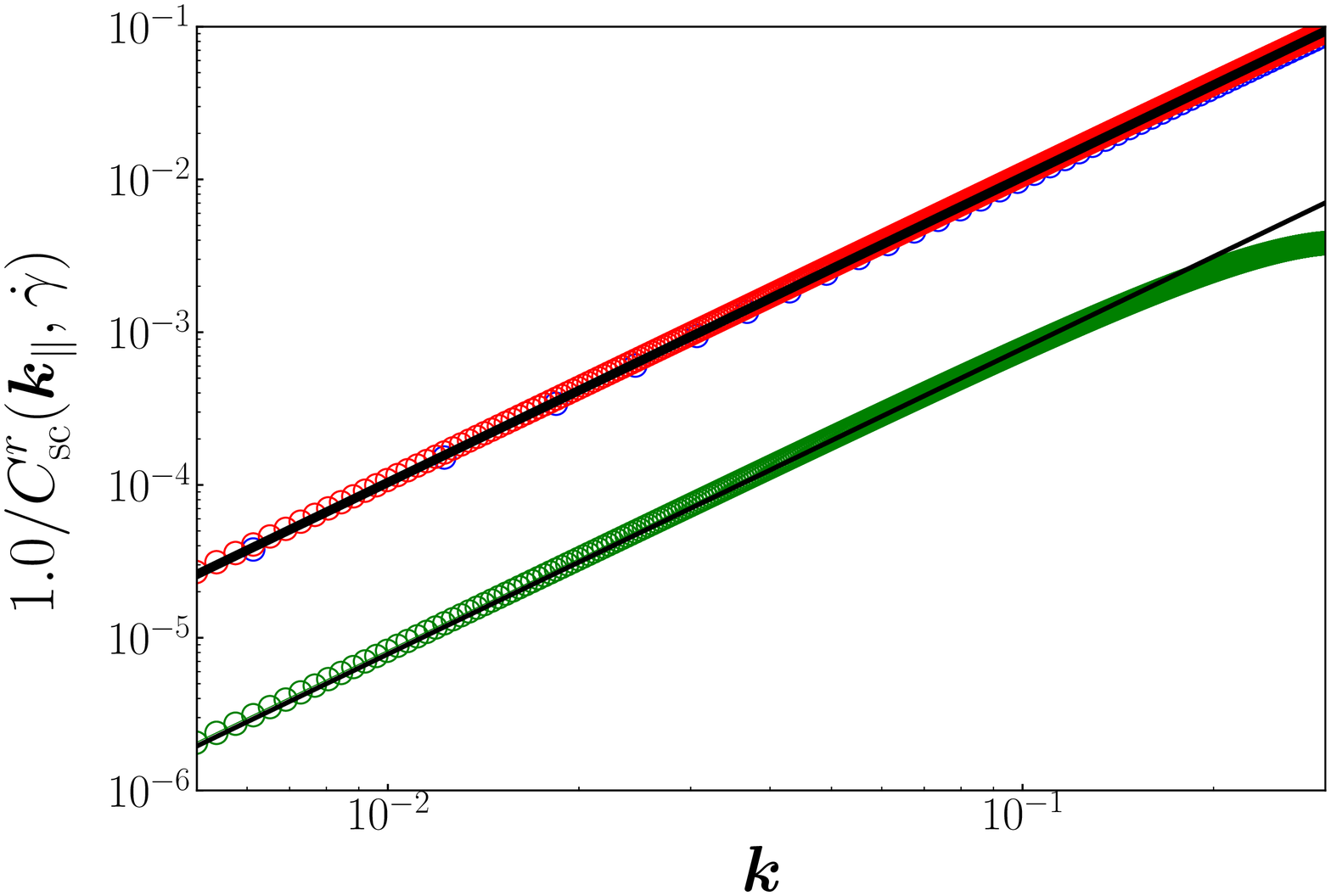}
\end{center}
\vspace{-0.5cm}
\caption{Left: correlation function in the disordered bulk. Red circle: $1.0/C^r_l(k_1,k_2=k_3=0;\dot{\gamma})-r$ vs. $k_1$. Blue circle: $1.0/C^r_l(k_1=0,k_2,k_3=0;\dot{\gamma})-r$ vs. $k_2$. Green circle: $1.0/C^r_l(k_1=k_2=k,k_3=0;\dot{\gamma})-r$ vs. $k$. The black solid curves are Eq.~(\ref{eq:sup:correlation function: c0=0: non-conserved}) for the corresponding wavenumber. Right: correlation function at the critical point. Red circle: $1.0/C_{\rm sc}^r(k_1,k_3=0;\dot{\gamma})$ vs. $k_1$. Blue circle: $1.0/C_{\rm sc}^r(k_1=0,k_3;\dot{\gamma})$ vs. $k_3$. Green circle: $1.0/C_{\rm sc}^r(k_1,k_3=0;\dot{\gamma})-k_1^2/\sqrt{r}$ vs. $k_1$. The black solid curves are Eq.~(\ref{eq:sup:correlation function: under shear c=0 model a: final}) for the corresponding wavenumber. The parameters are set to $r=T=\dot{\gamma}=1.0$, and $a_2^{\rm uv}=0.01$.  }
\label{fig: correlation function: A sr=1 r=1}
\end{figure}

\end{widetext}

\end{document}